\documentclass[twocolumn]{emulateapj}
\usepackage{graphicx}
\usepackage{natbib}
\newcommand{\eg}{{\it e.g.,}}
\newcommand{\ie}{{\it i.e.,}}

\begin{document}


\title{Kronoseismology: Using density waves in Saturn's C ring \\ to probe the planet's interior}
\author{M.M. Hedman$^*$ and P.D. Nicholson}
\affil{Center for Radiophysics and Space Research, Cornell University, Ithaca NY 14850 \\
$^*$Corresponding Author {\tt mmhedman@astro.cornell.edu}  \\
\\
{\em Proposed Running Head:} Kronoseismology with Saturn's rings. \\
\\}
\shorttitle{Kronoseismology with Saturn's rings.}

\begin{abstract}
Saturn's C ring contains multiple spiral patterns that appear to be density waves driven by periodic gravitational perturbations. In other parts of Saturn's rings, such waves are generated by Lindblad resonances with Saturn's various moons, but most of the wave-like C-ring features are not situated near any strong resonance with any known moon. Using stellar occultation data obtained by the Visual and Infrared Mapping Spectrometer (VIMS) onboard the Cassini spacecraft, we investigate the origin of six unidentified C-ring waves located between 80,900 and 87,200 km from Saturn's center. By measuring differences in the waves' phases among the different occultations, we are able to determine both the number of arms in each spiral pattern and the speeds at which these patterns rotate around the planet. We find that all six of these waves have between 2 and 4 arms and pattern speeds  between 1660$^\circ$/day and 1861$^\circ$/day. These speeds are too large to be attributed to any satellite resonance. Instead they are comparable to the predicted pattern speeds of waves generated by low-order normal-mode oscillations within the planet [Marley \& Porco 1993, Icarus 106, 508]. The precise pattern speeds associated with these waves should therefore provide strong constraints on Saturn's internal structure. Furthermore, we identify multiple waves with the same number of arms and very similar pattern speeds, indicating that multiple $m=3$ and $m=2$ sectoral ($l=m$) modes may exist within the planet. 

\medskip


\end{abstract}

\maketitle

\section{Introduction}

Spiral waves are patterns in Saturn's rings produced by periodic gravitational perturbations on the ring material. Many of these waves can be attributed to mean-motion resonances with Saturn's various moons. However, a number of spiral waves in the C ring do not fall near any known resonance with any moon. These waves were first noticed in the Voyager radio occultation data \citep{Rosen91}, but have also been observed in stellar occultations by the ultraviolet spectrometer onboard the Cassini spacecraft \citep{Colwell09, Baillie11}.  However, these previous studies were unable to identify the source of the perturbations responsible for generating these waves because they could not determine either the number of arms in the spirals or how fast the relevant patterns were rotating around the planet. Using occultation data from the Visual and Infrared Mapping Spectrometer (VIMS), we have now been able to determine both these quantities for six of these unidentified waves. These six waves have the right pattern speeds and symmetry properties to be produced by low-order normal mode oscillations within Saturn, as predicted by \citet{Marley90} and \citet{MarleyPorco93}. These waves should therefore  provide valuable new constraints on the planet's interior structure

Prior to describing our analysis of these waves, we first provide some background information about the relevant spiral waves in Section~\ref{background}. This includes a brief summary of the theory behind spiral waves (Section~\ref{approach}), a summary of the wave-like features in the C ring (Section~\ref{context}) and a description of the six waves that will be investigated here (Section~\ref{waves}). Section~\ref{methods} then describes how we can determine the number of arms and pattern speeds of these waves.  Section~\ref{data} describes the VIMS occultation data that will be used in this analysis. Section~\ref{rscnc}  illustrates how comparing multiple occultations can constrain the waves'  symmetry properties using a particularly informative set of occultations. Sections~\ref{phase} and~\ref{pattern} describe the wavelet-based techniques we employ to ascertain the number of arms and pattern speeds of these waves. Section~\ref{results} summarizes the results of our calculations. Finally, Section~\ref{discussion}  discusses the implications of our findings.

\section{Background}
\label{background}

\subsection{The theory of spiral waves}
\label{approach}

Spiral patterns in Saturn's rings are generated by various periodic perturbations on the ring material, and different types of perturbations produce different types of spiral waves. For example, periodic vertical forces generate warped structures known as bending waves, while periodic radial or azimuthal forces produce variations in the rings' surface density known as density waves.  The structure and dynamics of these spiral patterns are relatively well understood and good theoretical overviews of these phenomena are available  (e.g. Shu 1984). Hence, we will only briefly review the aspects of spiral waves that are most relevant to this analysis.  Note that the waves considered below all appear to be density waves, so we will focus on the dynamics of those structures here. \nocite{Shu84} 

A given spiral density wave consists of an integer number of arms that become more tightly wrapped with increasing distance from the radius of the exact resonance. This entire pattern rotates around the planet at a single pattern speed. The waves in Saturn's rings are so tightly wrapped that the opacity variations appear to be almost purely radial in both images and occultation profiles. Hence the waves appear as  periodic variations in ring brightness or opacity whose wavelength changes systematically with radius (see Figure~\ref{RCas106} below for some examples). However, because a density  wave is actually a rotating spiral pattern, the locations of the peaks and troughs in a given profile will depend on both the observed ring longitude and the orientation of the pattern during the measurement.  

The organized motions responsible for a spiral density waves are most efficiently generated by Lindblad resonances. At these resonances, the ring-particles' radial epicyclic frequency $\kappa$ is an integer multiple of the difference between the angular frequency of the perturbing potential $\Omega_p$ and the ring-particles' mean motion $n$. Hence, if there is a density wave driven at a given radial location in the rings, the most likely resonant perturbation frequencies will satisfy the following relationship:
\begin{equation}
m(n-\Omega_p)=\kappa
\label{eqone}
\label{patspeed1}
\end{equation}
where $m$ is any non-zero integer (\ie\ $m$ = ...-3, -2, -1, 1, 2, 3...).
Rewriting the resonant condition in terms of the local apsidal
precession rate, $\dot{\varpi} = n-\kappa$, we have the familiar
expression for a first-order Lindblad resonance:
\begin{equation}
(m-1)n + \dot{\varpi} = m \Omega_p.
\label{patspeed2}
\end{equation}
\noindent Since $\dot{\varpi} << n$, resonances with $m>0$ (known as inner Lindblad resonances or ILRs) have $\Omega_p \approx (m-1)n/m$, while those with $m<0$ (known as outer Lindblad resonances or OLRs) have $\Omega_p \approx (|m|+1)n/|m|$.

In a differentially-rotating, self-gravitating disk, the periodic perturbations at such a resonance give rise to a trailing spiral density wave that propagates away from the location of the exact resonance towards the location in the rings where $n = \Omega_p$. Thus, for a Keplerian disk like Saturn's rings (and assuming $\Omega_p>0$),  density waves will propagate outwards from an inner Lindblad resonance and inwards from an outer Lindblad resonance.
The pattern of surface density variations generated by such a wave gives rise to variations in the local optical depth with radius $r$, inertial longitude $\lambda$ and time $t$. For waves of small amplitude these variations may be written as:
\begin{equation}
\tau(r,\lambda,t) \simeq \tau_0 + \Delta\tau(r)\cos \phi(r,\lambda,t),
\label{eqtwo}
\end{equation}
where the wave's phase $\phi$ can be decomposed into a part that depends only on the observed longitude and time, and another that depends only on radius:
\begin{equation}
\phi(r,\lambda,t) \simeq |m|(\lambda - \Omega_p t)+\phi_r(r).
\end{equation}
Hence $|m|$ gives the number of arms in the wave pattern, while $\Omega_p$ is the angular rate at which it rotates around the planet. The wave's pattern speed therefore equals the angular frequency of the external perturbing force.  

At sufficiently large distances  from the resonance, the radius-dependent part of the phase is given by the following asymptotic expression
\begin{equation}
\phi_r(r) \simeq \left[3(m-1)+J_2\frac{21}{2}\left(\frac{r_S}{r_L}\right)^2\right]\frac{M_S(r-r_L)^2}{4\pi\sigma_0 r_L^4} + \phi_0,
\label{phir}
\end{equation}
where $M_S$ is the mass of Saturn, $J_2$ is a measure of the planet's oblateness, $r_S=60,330$~km, $r_L$ is the resonant radius where Eq.~(\ref{eqone}) is satisfied, $\sigma_0$ is the undisturbed surface mass density of the ring and $\phi_0$ is a constant. Note that for an outward-propagating wave from an  ILR both $m-1$ and $r-r_L$ are positive, while for an inward-propagating wave from an OLR $m-1$ and $r-r_L$ are both negative. Hence in both cases $d\phi/dr>0$ and a line of constant phase will have $dr/d\lambda < 0$, corresponding to a trailing wave pattern. Also, the wave's radial wavenumber $k$ derived from the above expression:
\begin{equation}
k(r) = \frac{d\phi}{dr} \simeq \left[3(m-1)+J_2\frac{21}{2}\left(\frac{r_S}{r_L}\right)^2\right]\frac{M_S(r-r_L)}{2\pi\sigma_0 r_L^4},
\label{kr}
\end{equation}
\noindent increases linearly with distance from $r_L$. 

So long as the background opacity of the ring $\tau_o$ does not vary too much with radius, the phase parameter $\phi$ can be computed directly from the local opacity variations (\eg\ opacity maxima occur where $\phi \simeq 0$ and opacity minima occur where $\phi \simeq 180^\circ$). Thus $\phi(r, \lambda)$ can be calculated from any given radial profile across the ring. Indeed, for identifiable waves the trends in $\phi$ with radius can be used to estimate parameters like the ring's surface mass density \citep{Tiscareno07}. However, for our efforts to determine the pattern speeds and $m$-values of unidentified waves, it is more useful to consider the difference in phase parameters between occultations observed at different times and longitudes. For these phase differences $\delta \phi$, the radius-dependent term $\phi_r$ should cancel out, leaving terms that only depend on the known differences in the observation times ($\delta t$) and observed longitudes $(\delta \lambda)$, and the unknown parameters $|m|$ and $\Omega_p$:
\begin{equation}
\delta \phi=|m|(\delta \lambda-\Omega_p\delta t).
\label{dphi}
\end{equation}
 We can therefore compare the observed phase differences $\delta \phi$ with those predicted for various combinations of $m$ and $\Omega_p$,  and thereby determine which pattern speed and $m$-number is most  consistent with the observed phase differences.

\subsection{Spiral waves in the C ring}
\label{context}

Surveys of Saturn's C-ring by \citet{Rosen91}, \citet{Colwell09} and \citet{Baillie11} have found a total of 27 wave-like structures in the C-ring. Only five of these features have been identified with specific satellite resonances. The most prominent feature is a bending wave generated by an unusual Titan -1:0 nodal resonance \citep{RL88},  and  \citet{Rosen91} identified one density wave produced by the Mimas 4:1 ILR. More recently, the density wave produced by the Atlas 2:1 ILR has been detected, and there are hints of density waves at the locations of the Mimas 6:2 and Pandora 4:1 resonances \citep{Colwell09, Baillie11}.  More than half of the remaining, unidentified waves appear to be inward-propagating (\ie\ their radial wavenumber increases inwards). Such behavior is inconsistent with density waves generated by resonances with Saturn's moons. Since all the moons orbit outside the rings, the perturbation frequency associated with any moon's gravitational perturbations will be smaller  than the local orbital frequency anywhere in the C ring. Hence Lindblad resonances in the C ring with any moon should be ILRs, which should generate outward-propagating waves. The inward-propagating waves must therefore be produced by some other mechanism.

One possible explanation for these inward-propagating waves is that they are bending waves instead of density waves.  Unlike density waves, bending waves typically propagate away from the location in the rings where $\Omega_p=n$ \citep{Shu84}. Hence vertical resonances where the perturbation frequency is less than the local orbital frequency (the so-called inner Vertical resonances or IVRs), like those generated by several of Saturn's moons, should  propagate inwards. Furthermore, the very low ring opening angle in the Voyager radio experiment ($B = 5.6^\circ$) meant that the vertical warps in the bending wave could lead to significant optical depth variations along the observed line of sight. However, most of these inward-propagating waves are also clearly visible in stellar occultations observed by Cassini at high ring opening angles, where optical depth variations from bending waves would be very subtle\footnote{One exception is the wave designated ``j'' in \citet{Rosen91}, which is in fact invisible in all VIMS occultation profiles obtained to date, but is seen in one or more UVIS stellar occultations observed at very low opening angles \citep{Baillie11}.}. Furthermore, there are no strong candidate IVRs due to known (or even hypothesized) satellites at the desired locations. Thus this is not the currently-favored interpretation for most of these inward-propagating features.

An alternative explanation of these waves is that they are not generated by resonances with Saturn's moons, but instead are produced by resonances with normal-mode oscillations within Saturn itself. \citet{Stevenson82} first suggested that oscillations within the planet could produce identifiable structures in the rings. Later, \citet{Marley90,   Marley91} and \citet{MarleyPorco93}  demonstrated that acoustic modes in the planet's interior could indeed give rise to gravitational perturbations with pattern speeds that are sufficiently fast to generate OLRs and inward-propagating waves in the C ring. The latter authors also found that the predicted pattern speeds of low-degree $f$-modes  were consistent with some of  the observed wave locations, but uncertainties in Saturn's interior structure and in the theory of normal modes meant that the resonance radii could only be predicted to within $\sim500$~km. 

\subsection{The six waves examined in this study}
\label{waves}


\begin{figure*}[tbp]
\centerline{\resizebox{6in}{!}{\includegraphics{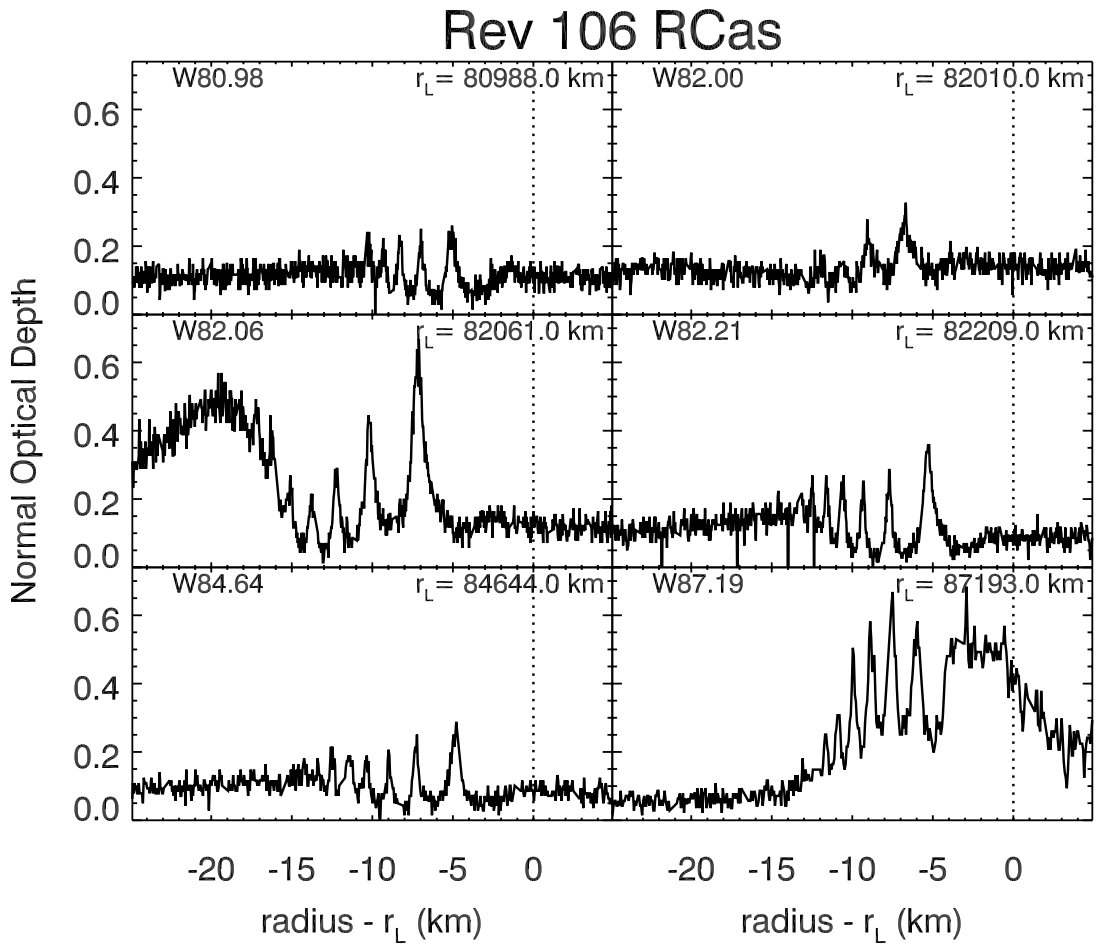}}}
\caption{VIMS occultation profiles of the six waves examined in this analysis. Each panel shows the ring's normal optical depth  versus ring radius,  which is measured in kilometers from the inferred resonance location $r_L$. (For the innermost five waves, the $r_L$ value comes from Bailli\'e {\em et al.} 2011, while for W87.19 the resonance position has been adjusted to match the best-fit pattern speed of this wave, see Seciton~\ref{results}.) The specific profiles shown here come from an occultation by the rings of the star R Cassiopaea, which provides our highest-resolution profiles of these waves. The raw data numbers were converted to transmission estimates by normalizing the stellar signal to unity in the middle of the Maxwell Gap (87,375-87,425 km), and then translated to normal optical depth values using the standard formula, assuming the  elevation angle of the star is 56.04$^\circ$ above the ringplane. }
\label{RCas106}
\end{figure*}
   
\begin{table}[tbp]
\caption{Summary of waves investigated here, including designations by
various authors.}
\medskip
\label{wavetab}
\centerline{\begin{tabular}{|c|c|c|c|c|c|c|} \hline
Wave & Rosen {\it et al.} & Colwell {\it et al.} & Baillie {\it et al.} \\
 Location$^a$ & (1991)$^b$ & (2009b) & (2011) \\\hline 
80988 km &  e &  W80.98 & 13  \\
82010 km &      & W82.00 & 15  \\
82061 km & f   & W82.06 & 16 \\
82209 km & g & W82.21 & 17  \\
84644 km & i   & W84.64 & 19 \\
87189 km &     & W87.19 & 32 \\ \hline
\end{tabular}}
\medskip
$^a$ Inferred resonance location obtained by fitting Eq.~\ref{kr} to UVIS occultation data, see  Table 7 of \citet{Baillie11}. \\
$^b$ See Figures 7 and 8 in \citet{Rosen91}.
\end{table}


This analysis will focus on six of the unidentified, inward-propagating waves. As summarized in {Table~\ref{wavetab}}, these waves have been designated in various ways by different authors. Here we will use the \citet{Colwell09} nomenclature, which identifies each wave with a number giving the radial location of the wave in thousands of kilometers. Figure~\ref{RCas106} shows profiles of all six waves derived from the highest resolution VIMS occultation to date (see below for how these data were processed). Each wave appears as a periodic variation in the ring's opacity, with a wavelength that varies with distance from the planet. More specifically, all these waves have wavelengths which decrease inwards, which suggests that they are either bending waves driven by IVRs or density waves driven by OLRs. Since the visibility of these waves does not appear to vary with spacecraft elevation angle, it is unlikely that any of these features are bending waves. Indeed, their pattern speeds turn out to be consistent with those expected for OLRs. Hence for the remainder of this discussion we will anticipate our final result by referring to these features as density waves.

All six waves are located in the central part of the C ring, between radii of 80900 and 87200~km. This is a region of gently undulating structure, with an average normal optical depth (\ie\ the optical depth the ring would have if the line of sight was exactly perpendicular to the ringplane) of about 0.10. All but the outermost wave are located between the `plateau' features designated P4 and P5 by \citet{Colwell09}. While four of the waves fall in otherwise unexceptional locations, two fall either on top of (W87.19) or immediately adjacent to (W82.06) local maxima in optical depth.  

The strongest known eccentric resonances in this region are the two 4th-order Enceladus 5:1 resonances at 82538 and 82477~km, the 3rd-order Janus and Epimetheus 5:2 resonances at $\sim82780$~km and $\sim82950$~km, the Pan 2:1 ILR at 85105~km, and the Atlas 2:1 ILR at 87647~km. Only the last of these is known to be associated with an observable wave. None of these resonances is located within 100~km of any of the unidentified waves. There are two very weak vertical resonances in the vicinity of our target waves: the Enceladus 5:1($e^2i$) resonance at 80967~km (located 21 km interior to the W80.98 wave), and the Janus 5:2($ei$) resonance at 82096-82107~km ($\sim$40 km exterior to the W82.06 wave). However, as mentioned above, the visibility of these waves at large ring opening angles argues against these features being bending waves, and furthermore many other comparable (or stronger) vertical resonances  do not produce visible bending waves in the rings. Hence we consider these rough alignments to be coincidental, and conclude that none of these six waves can be attributed to a resonance with any of Saturn's moons.


These six waves have the strongest opacity variations and the longest wavelengths of any of the unidentified C-ring waves. We focus exclusively on these waves because they provide the best opportunities for determining unambiguous pattern speeds. The strong opacity variations mean that we can clearly detect the maxima and minima, which makes the relevant phase parameters easier to determine. Furthermore, the relatively long wavelengths of these waves should minimize our sensitivity to small errors in the occultation geometry. Errors in Cassini's trajectory reconstruction can cause the occultation profiles to shift slightly in radius, which can in turn create problems when comparing data from different occultations. For example, if one profile was shifted relative to the other by half a wavelength, then the phase difference between the waves might appear to be zero when in reality it is 180$^\circ$. However, this is unlikely to occur with any of these six waves, because they all have maximum wavelengths exceeding 2 kilometers, which is considerably larger than the sub-kilometer uncertainties in our reconstructions of occultation geometry (see below).

\section{Methods}
\label{methods}

\subsection{VIMS Observations}
\label{data}

The VIMS instrument is described in detail in \citet{Brown04}. While VIMS is typically used to obtain spatially-resolved spectra of a scene, it can also operate in an ``occultation mode'' where the short-wavelength VIS channel is turned off, while the longer-wavelength IR channel stares at a single pixel targeted at a star and obtains a series of rapidly-sampled near-infrared stellar spectra at 31 wavelengths between 0.85 and 5.0 microns. As the star moves behind the rings, its apparent brightness varies due to variations in the ring's opacity. Note that the response of the detector is highly linear, so after a constant instrumental background is removed from each spectral channel, the data numbers returned by the instrument are proportional to the incident flux. In order to avoid contamination from sunlight scattered by the rings, we focus exclusively on data from one spectral channel covering the range 2.87-3.00 microns, where water ice is strongly absorbing.  The rings are sufficiently dark at these wavelengths that ringshine is negligible, and the measured signal is directly proportional to the transmission through the rings $T$. We can therefore easily translate the raw data numbers into the slant optical depth along the line of sight through the rings $\tau=-\ln(T)$, or the normal optical depth of the rings $\tau_n=\tau\sin B$ ($B$=elevation angle of the star above the rings).

A precise time stamp is appended to each spectrum, facilitating the geometry reconstruction. Using the appropriate SPICE kernels, we can compute the radius where the starlight pierced the ringplane for each sample in the given occultation. This calculation accounts for the light travel time from the rings to Cassini, and uses stellar positions taken from the Hipparcos catalog, corrected for parallax at Saturn.  The positions of sharp edges of gaps and ringlets in the Cassini Division and C ring demonstrate that the resulting reconstructed geometry for each occultation is accurate to within   a kilometer \citep{Nicholson11, French11}. French {\it et al.} (in prep) have used a sub-set of these sharp features  to make small corrections to the spacecraft's position during these occultations. These corrections not only reduce the  dispersion in the radial position estimates of sharp edges to $\sim150$ meters, but also yield more consistent estimates of the density wave phases (see below).

Between 2005 and 2009, VIMS obtained a total of 27 occultation cuts through the C ring with sufficient signal-to-noise and spatial resolution to discern the relevant waves. Table~\ref{occtab1} provides a list of these occultations, identifying the star observed, the ``Rev number'' (Cassini's orbit around Saturn) when the data were obtained, and whether the cut was obtained during ingress or egress. The table also provides the ephemeris time and observed inertial longitude when the star passed behind each of the six waves discussed below. Blank entries in this table correspond to cases where the occultation did not cover the particular wave or when a data gap corrupted the relevant profile, and thus were excluded from this analysis.

\subsection{Initial examination of the waves using RS Cancri occultations}
\label{rscnc}

\begin{figure*}[tbp]
\centerline{\resizebox{6in}{!}{\includegraphics{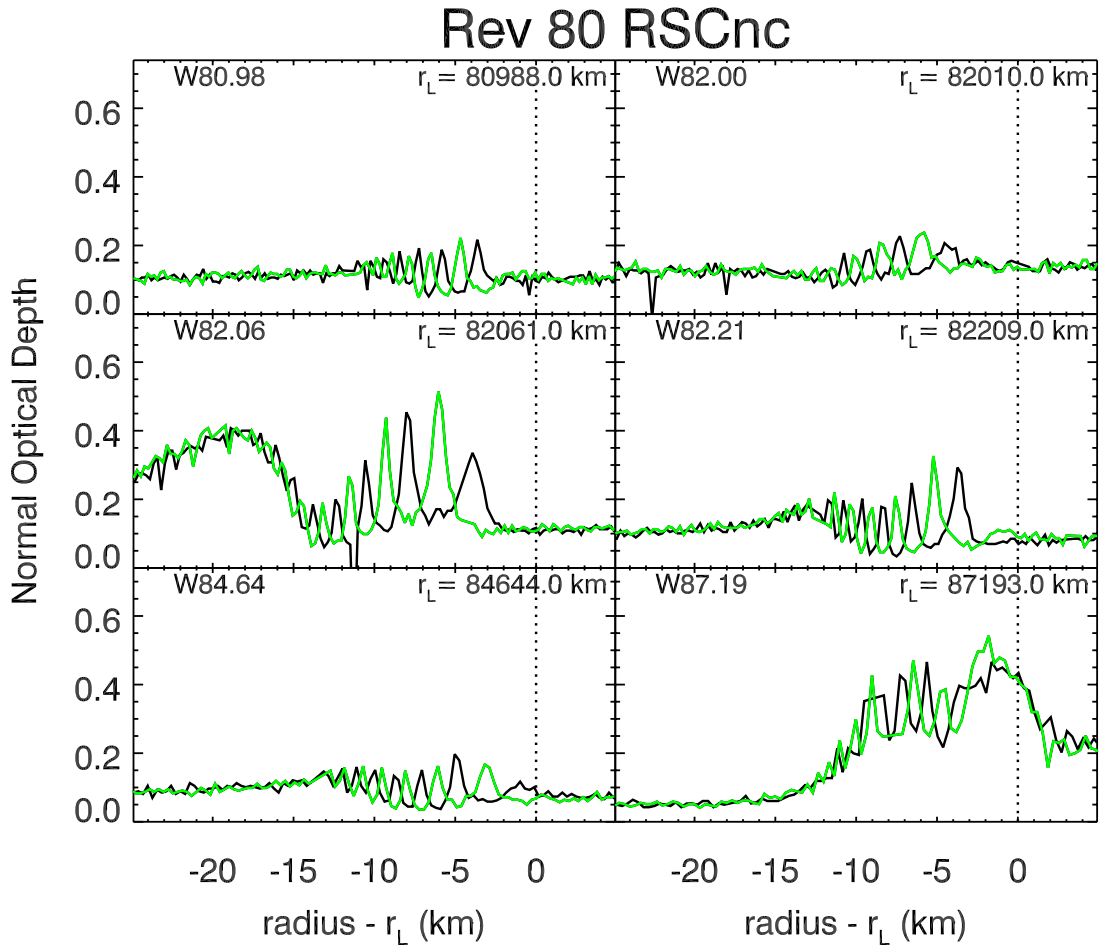}}}
\caption{Profiles of the relevant waves obtained during the occultation of the star RS Cancri on Rev 80. The black profiles were obtained during ingress, while the green profiles were obtained on egress. The normal optical depth values assume the star's elevation angle above the rings is 29.96$^\circ$. Note that the phase differences between the waves seen in ingress and egress are the same for the three waves found around 82,000 km. This suggests that all these waves have the same $m$-numbers.}
\label{RSCnc80}
\end{figure*}

\begin{figure*}[tbp]
\centerline{\resizebox{6in}{!}{\includegraphics{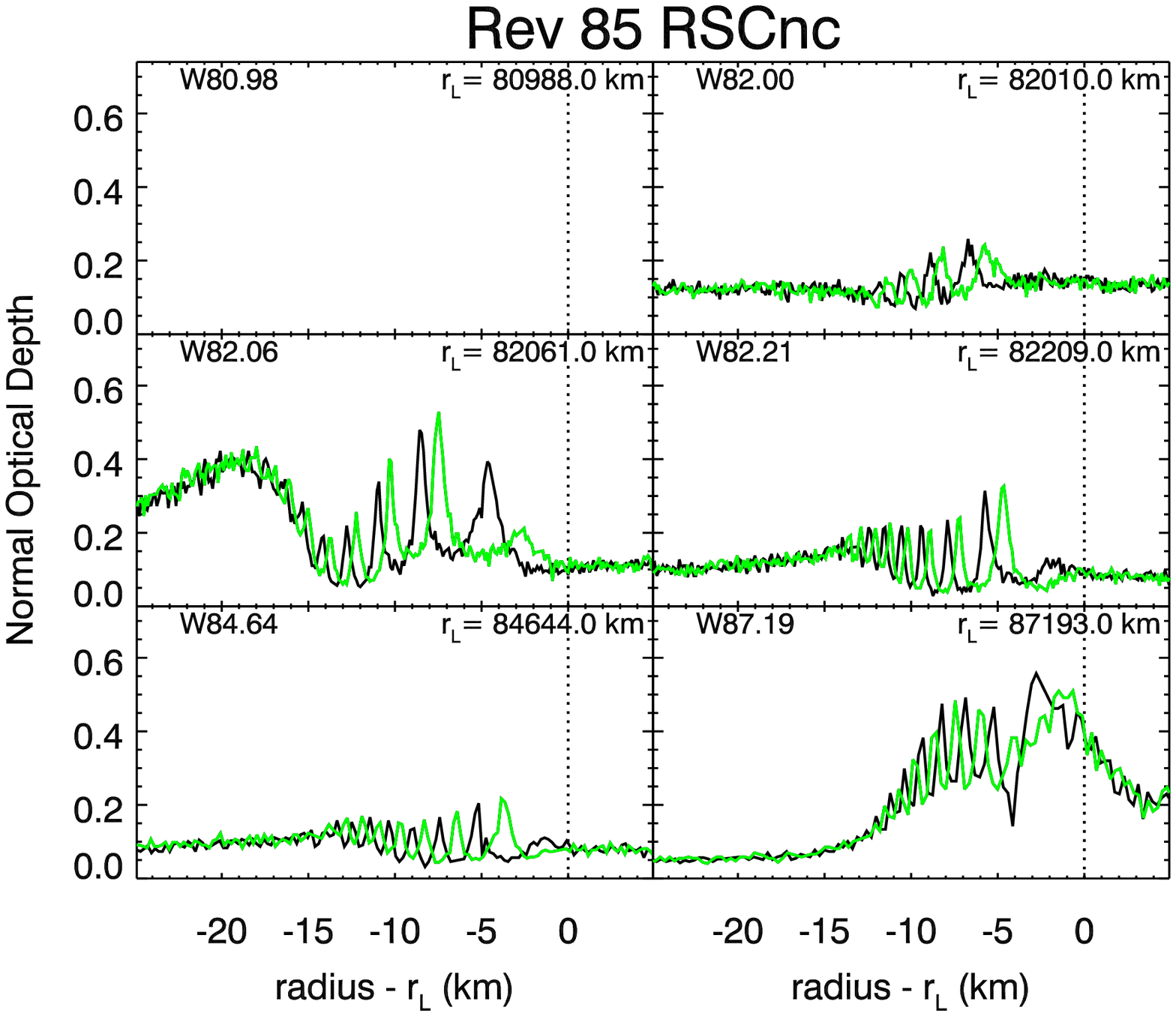}}}
\caption{Profiles of the relevant waves obtained during the occultation of the star RS Cancri on Rev 85. The black profiles were obtained during ingress, while the green profiles were obtained on egress. Note that the phase differences between the waves seen in ingress and egress are again the same for the three waves found around 82,000 km. This implies that all these waves have the same $m$-numbers. The two waves found outside 84,000 km also show similar phase differences, indicating that they may have same $m$-number as each other.}
\label{RSCnc85}
\end{figure*}

Before describing the procedures we will use to determine the $m$-numbers and pattern speeds of the waves from these occultation data, let us first examine some particularly informative occultations of the star RS Cancri. Unlike the other occultations considered in this analysis, these were chord occultations where the star cut obliquely through the rings. In such an occultation, the star passes through the same radial range at two very different longitudes as the track enters and then leaves the rings. Comparisons of the wave profiles derived from the ingress and egress parts of these occultations thus provide illustrative examples of how occultation data can constrain wave parameters. Furthermore, even simple visual inspections of these wave profiles 
indicates that several of the unidentified waves have very similar pattern speeds and likely the same $m$-numbers.

{Figures~\ref{RSCnc80}} and {\ref{RSCnc85}} show the relevant occultation profiles from the RS Cancri occultations observed during Revs 80 and 85 (the Rev 87 RS Cancri occultation only probed waves exterior to 84,000 km and is not shown here). Of particular interest are the three waves W82.00, W82.06 and W82.20, which are clustered within a region a couple of hundred kilometers wide around 82,100 km.  In the Rev 80 RS Cancri data the ingress and egress profiles for each of these waves are almost perfectly anti-correlated, with peaks in one profile corresponding to dips in the other and vice versa. This indicates that the phase difference between these two occultation cuts is close to 180$^\circ$ for all three of these waves. This finding alone might be interpreted as a coincidence, but if we turn our attention to the Rev 85 data, we again find suspicious similarities among these three waves. For all three of the waves, the sharp peaks in the egress data occur about one-third of a cycle exterior to the sharp peaks in the ingress data. This again suggests that the phase difference between the ingress and egress cuts (in this case $\sim240^\circ$) is nearly the same for all three of these waves.

This result is significant because it implies that all three of these waves almost certainly have the same $m$-number. The waves W82.00, W82.06, and W82.21 are close together in the rings. Hence, for each of these occultations, the ingress cuts for these three waves occurred at nearly the same longitude and time, and similarly the egress cuts are grouped closely in longitude and time (see Table~\ref{occtab1}). This means the difference in the observed longitudes ($\delta\lambda$) and the observed times ($\delta t$) between ingress and egress are almost identical for the three waves. Since the phase difference between the waves is given by $\delta \phi=|m|(\delta \lambda-\Omega_P \delta t)$, the easiest way for all three waves to have the same phase difference is for them to have the same $m$-number (which also implies, via Eq.~\ref{patspeed2}, nearly identical pattern speeds). Our full analysis of these waves confirms that this is indeed the case.

Being able to determine that W82.00, W82.06 and W82.21 have the same $m$-number from simple inspection of the RS Cancri profiles is useful because this result does not require the geometry of the occultations to be extremely accurate. The most likely errors in the occultation geometry correspond to shifts in the spacecraft's position along its orbit, which would cause the ingress and egress occultation traces to shift in opposite directions in radius.  Such errors will tend to shift the ingress profiles outward or inward relative to the egress curves by the same amount for all three waves. Although the ingress-egress phase differences would then change, as long as the wavelengths are similar this would not change the result that the phase differences between the ingress and egress cuts are nearly the same for all three waves.


\subsection{Computing phase differences between waves}
\label{phase}

While simple inspection of the RS Cancri profiles suggests that several of the waves have similar pattern speeds and identical $m$-numbers, we need quantitative measurements of the phase differences between different occultation cuts to ascertain the actual values of $m$ or $\Omega_p$ for these waves. Fortunately, we can estimate the phase difference between any two occultation cuts through a given wave using wavelet transforms. A wavelet is basically a localized Fourier transform that can effectively cope with the rapid wavelength changes in typical spiral waves. Indeed, wavelet techniques have already been used to study density waves in order to obtain information about how the radial wavelength varies with distance from the resonance, which can constrain the rings' surface mass density \citep{Tiscareno07, Colwell09cd, Baillie11}. However, our analysis is different in that it focuses on estimating the {\em phase difference} between two wave profiles rather than the wavelength or amplitude of the wave.

\nocite{TC98}
We perform our wavelet analysis on the raw signal profiles instead of the derived optical depth profiles shown in Figures~\ref{RCas106}-~\ref{RSCnc85}. In practice, this choice has little influence on the derived phase differences (the conversion from $T$ to $\tau$ just introduces an overall constant phase shift of $180^\circ$), but it does eliminate any possible complications that could arise due to the uncertainties in the unocculted star signal.  Before applying the wavelet transform to the relevant occultation data, we first interpolate the data onto a uniform array of radii with a radial spacing of 50 m. A small part of this re-sampled profile centered on each wave (see Table~\ref{waverscnc} or ~\ref{wavetab2} for the exact ranges) is then fed into the publicly-available IDL routine {\tt wavelet} (see Torrence and Compo 1998, the default Morlet mother wavelet with $\omega_0=6$ is used throughout this analysis). The resulting wavelet is a two-dimensional array of complex numbers as a function of radius and radial wavenumber $\mathcal{W}(r, k)$. (Note we use the wavenumber $k$ here instead of the wavelength $\lambda$ in order to avoid any possible confusion with the radiation's wavelength or the longitudinal coordinate.) Let us denote the real and imaginary parts of the wavelet as $\mathcal{W}_R$ and $\mathcal{W}_I$, respectively. We can then define the wavelet power as
$\mathcal{P}(r,k)=\mathcal{W}_R^2+\mathcal{W}_I^2$ and the wavelet
phase as $\varphi(r, k)=\tan^{-1}(\mathcal{W}_I, \mathcal{W}_R)$.

\begin{figure*}
\centerline{\resizebox{6in}{!}{\includegraphics{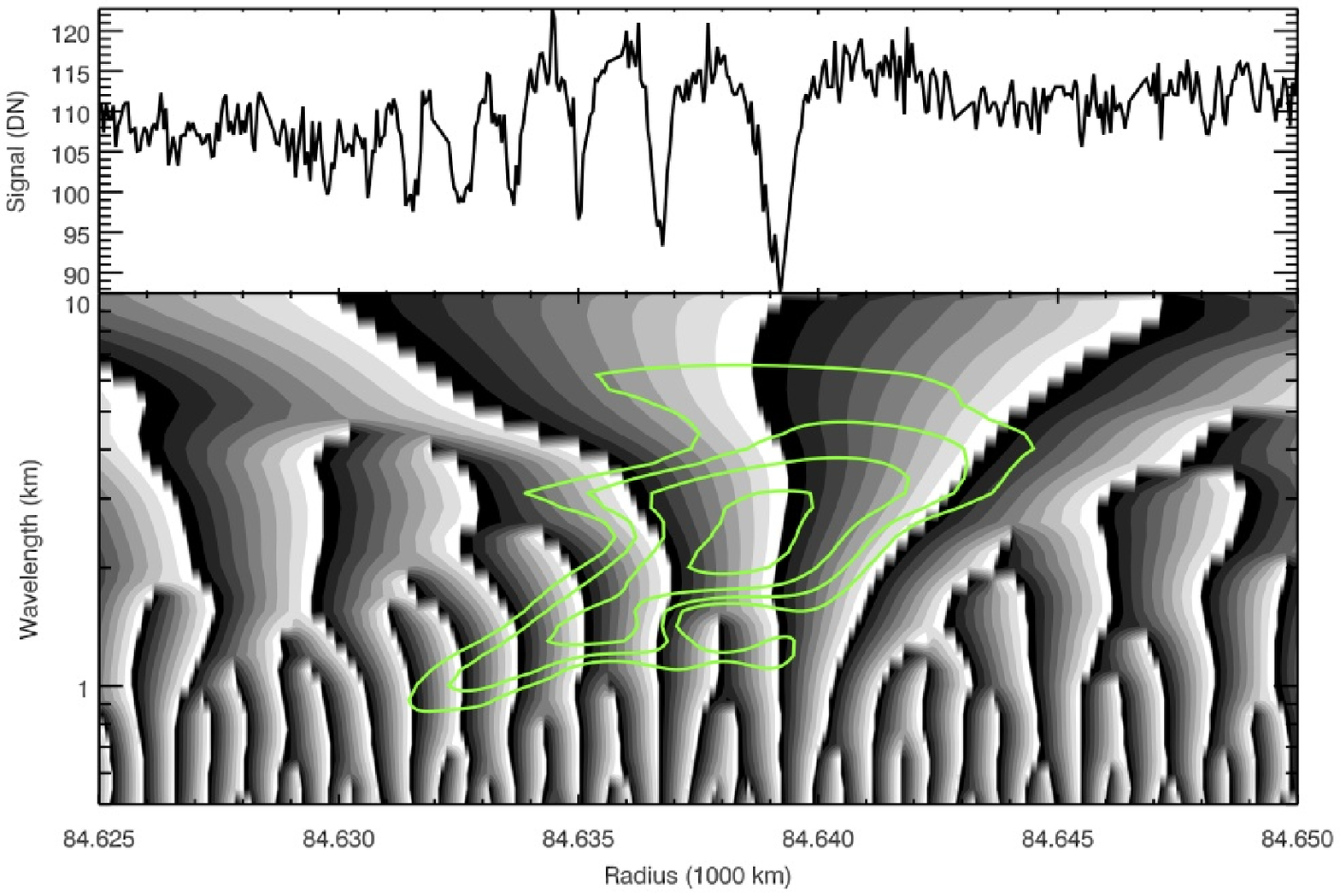}}}
\caption{Plot showing the wavelet amplitude and phase derived from the W84.64 wave observed by the Rev 106 RCas occultation. The top panel shows the occultation profile (in raw Data Numbers, which is proportional to transmission) as a function of radius. The bottom panel shows the wavelet phase and power as functions of radius and spatial wavelength. The wavelet phase is indicated by greyscale levels (black$=-180^\circ$, white$=+180^\circ$) while the overlaid green contours are levels of constant wavelet power. The peak wavelet power follows a diagonal ridge that corresponds to the wave's increasing wavelength with radius. Note that where the wavelet power is strong, the contours of wavelet phase are nearly vertical and  correspond to the expected phase of the wave (\eg\  the phase wraps from -180$^\circ$ to $180^\circ$ at locations corresponding to sharp minima in the profile). }
\label{wavedisp}
\end{figure*}

{Figure~\ref{wavedisp}} illustrates how the wavelet phase and power vary with position and wavelength across the wave. As expected there is a diagonal ridge in the wavelet power that tracks the observed trends in the wave's wavelength. (Recall that for a wave generated at an OLR, the radial wavelength decreases as the wave propagates inwards.) Furthermore, we can observe that where the wave is strong and the wavelet power is high, the contours of constant phase are nearly vertical, so a well-defined phase can be ascribed to every radius in the wave. The values of the phase in this region are also reasonable given the profile, being near $\pm180^\circ$ at local minima and around 0 at local maxima.\footnote{The wavelet phase also increases with increasing radius, consistent with the expected trend in $\phi(r)$ for a density wave.} Thus, for such waves we can reduce the two-dimensional wavelet to estimates of the wave power and phase as a function of radius by appropriately averaging over a range of spatial wavenumbers.

In order to filter out large scale background variations in the rings' opacity while still capturing most of the wave's power, we include a limited range of wavenumbers in these averages. Specifically we exclude all wavenumbers less than $k_1=2\pi/(5$ km) and all wavenumbers greater than $k_2=2\pi/(0.1$ km). The effective power of the wave at a given radius is therefore defined to be:
\begin{equation}
P_{\rm eff}(r)=\mathcal{N}\sum_{k=k_1}^{k_2}\mathcal{P}(r,k),
\end{equation}
where $\mathcal{N}$ is a normalization constant. Note that since we only are interested in relative power for this analysis, we choose $\mathcal{N}$ so that $P_{\rm eff}$ is equal to unity at its peak value. Also note that this sum is done over a series of logarithmically-spaced values of $k$.

In order to derive an effective phase for the wave at each radius, we first compute an effective average real part and imaginary part of the wavelet:
\begin{equation}
W_{R,I}(r)=\frac{\sum \mathcal{W}_{R,I}(r,k)\mathcal{P}(r,k)}
{\sum\mathcal{P}(r,k)}
\end{equation}
where the sums are again over all values of $k$ between $k_1$ and $k_2$. Note these averages are weighted by the power, so that the regions with the strongest signals dominate the averages. From these average wavelet components, we can compute the average phase at each radius:
\begin{equation}
\phi(r)=\tan^{-1}({W}_I,{W}_R)
\end{equation}
Note by computing the average of components $\mathcal{W}_I$ and $\mathcal{W}_R$ instead of averaging the local phases $\varphi(r,k)$, we avoid any difficulties involved in averaging a cyclic quantity. The phase difference between two occultations should then simply be the difference in the two values of $\phi(r)$.

\begin{figure}
\centerline{\resizebox{3in}{!}{\includegraphics{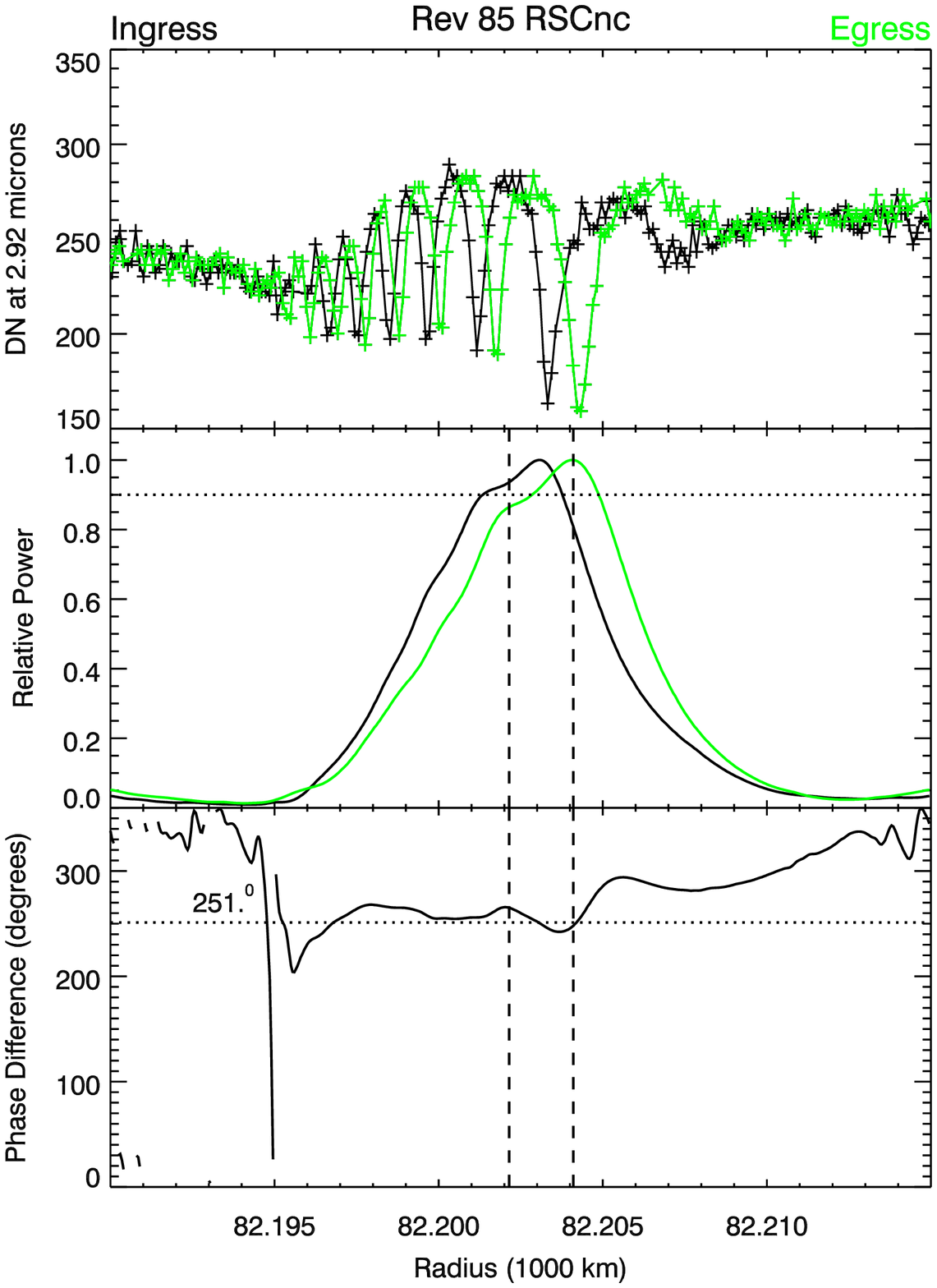}}}
\caption{Results of the wavelet calculations of the phase difference in wave W82.21 between the ingress and egress cuts from the Rev 85 RSCnc occultation. The top panel shows the two occultation profiles, while the middle panel shows the integrated wave power $P_{\rm eff}(r)$ between wavenumbers of $2\pi/(5$ km) and $2\pi/(0.1$ km). The bottom panel shows the phase difference $\delta \phi(r)=\phi_{egress}(r)-\phi_{ingress}(r)$ between these two cuts (see text for explanations of the dashed and dotted lines). Note that  the average phase difference is computed using only the data where the average $P_{\rm eff}$ of the two profiles is above 0.9. The average phase difference is near $240^\circ$, which is consistent with the offset between the ingress and egress wave profiles noted in Fig.~\ref{RSCnc85}.}
\label{wavefig1}
\end{figure}

\begin{figure}
\centerline{\resizebox{3.1in}{!}{\includegraphics{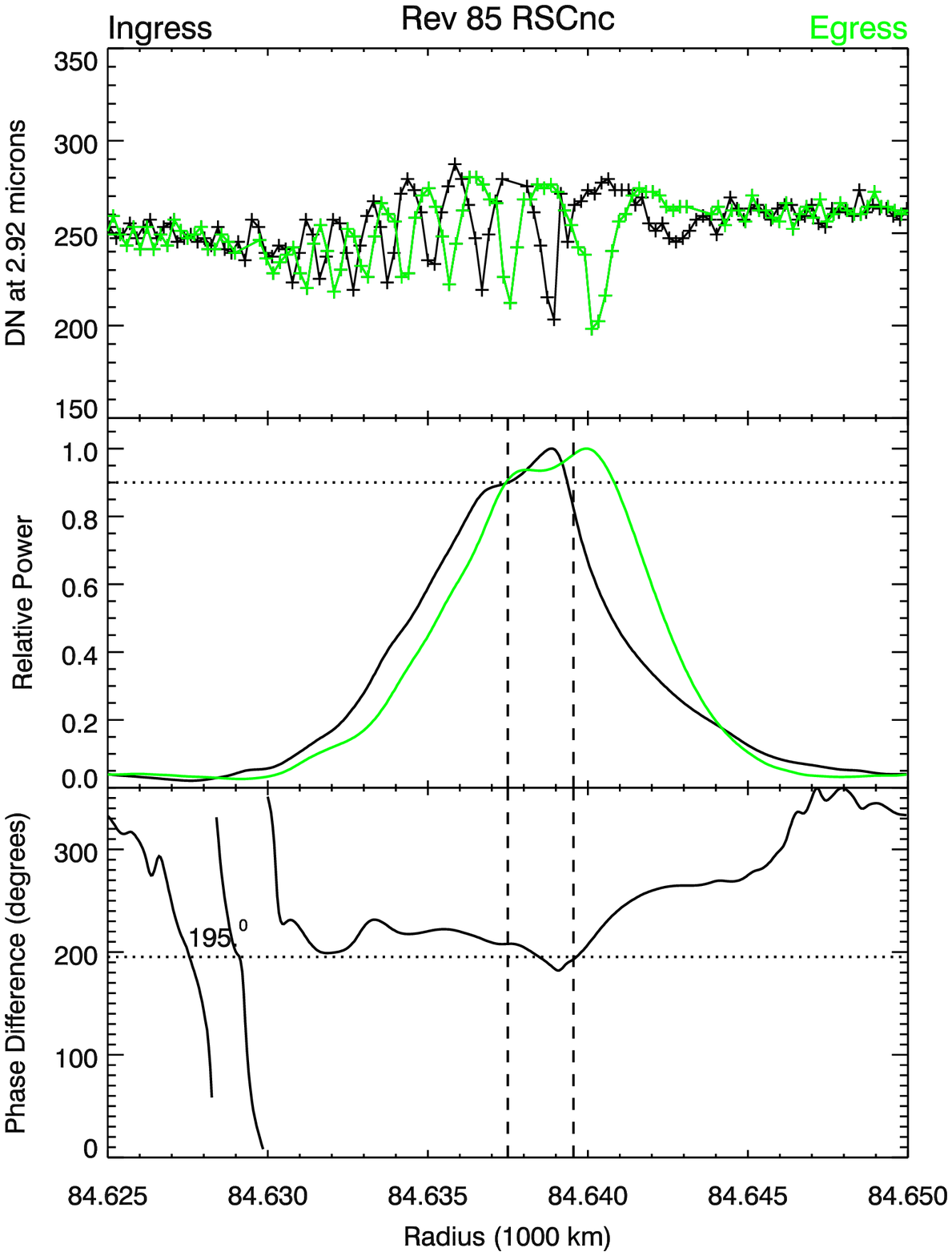}}}
\caption{Results of the wavelet calculations of the phase difference in wave W84.64 between the ingress and egress cuts from the Rev 85 RSCancri occultation, following the same layout as in Fig.~\ref{wavefig1}. In this case, the average phase difference is near $180^\circ$, which is again consistent with the ingress and egress wave profiles.}
\label{wavefig2}
\end{figure}

{Figures~\ref{wavefig1}} and {\ref{wavefig2}} illustrate the results of these calculations for two of the waves seen in the RS Cancri occultation from Rev 85. As expected, there is a peak in the wave power $P_{\rm eff}$ located near the center of the wave. Also, the difference in the wave phase between the egress and ingress cuts is relatively constant in the region where the wave signal is evident, as desired.  Furthermore, the numerical values of these phase differences are consistent with the observed profiles. For the W84.64 wave in Figure~\ref{wavefig2}, the phase difference is around 180$^\circ$, which is what one would expect given that the peaks in the ingress profile occur at the same location as the dips in the egress profile and vice versa. Similarly, the W82.21 data in Figure~\ref{wavefig1} yield a phase difference of around $240^\circ$, which is consistent with the dips in the egress profile always lying about $1/3$ of a cycle exterior to those in the ingress profile. This gives us some confidence that our rather simple approach can extract useful phase information from these wave profiles.

For the purposes of this analysis, we require a single estimate of the average phase difference for any pair of occultation cuts. We estimate this parameter as the weighted average of the phase differences $\delta\phi(r)$, with each of the individual values weighted by the average of the two $P_{\rm eff}$ curves in order to ensure that regions with high signal contribute most to the final estimate. In order to further reduce the possibility of contamination in the phase difference estimate, we only consider regions where the average power of the two waves is more than 0.9 (this threshold is indicated by a dotted line in Figures~\ref{wavefig1} and~\ref{wavefig2}, and the averaging region is demarcated by vertical dashed lines). Due to the weighting, the resulting estimates of the phase differences are not particularly sensitive to the exact value of this threshold. In addition to computing the weighted average phase difference, $\delta \phi$ between the two cuts (which is shown as a dotted line in the bottom panel of Figures~\ref{wavefig1} and~\ref{wavefig2}), we also compute the standard deviation of the phase difference values in the selected region $\sigma_\phi$, which quantifies the reliability of the phase estimate. Note, however, that $\sigma_\phi$ will underestimate the uncertainty in the $\delta \phi$ because it does not include the effects of uncertainties in the reconstructed geometry. Such geometric uncertainties are difficult to quantify {\it a priori}, and will be considered in more detail below.

To recap this procedure, the steps are: (1) Compute the wavelet transform, $\mathcal{W}(r,k)$ for each observation and the corresponding power spectrum, $\mathcal{P}(r,k)$. (2) Compute weighted averages for $W_R(r)$ and $W_I(r)$, and thence the average phase $\phi(r)$. (3) Compute the average radial power profile, $P_{\rm eff}(r)$. (4) Compute the average phase difference between two wave profiles $\delta\phi$, weighted by the average of the two power profiles.

\subsection{Using phase differences to constrain $m$-numbers and pattern speeds}
\label{pattern}

\begin{table*}[tbp]
\caption{Phase differences between ingress and egress cuts for the RS Cancri occultations}
\label{waverscnc}
\medskip
\centerline{\begin{tabular}{|c|c|c|c|c|c|c|} \hline
Wave & Resonant & Region  & Rev 80  & Rev 85  & Rev 87  & Possible  \\
 & Location$^a$ & Considered$^b$ & $\delta \phi$  &  $\delta \phi$   & $\delta \phi$ & $m$-values$^c$ \\ \hline
W80.98 & 80988 km &  80970-80995 km &  141.6$^\circ$ & --- & --- &  -4, -3, +6 \\
W82.00 & 82010 km & 81992-82012 km & 148.5$^\circ$ & 250.6$^\circ$ & --- & -3, +6\\
W82.06 & 82061 km & 82040-82065 km & 148.1$^\circ$ & 249.1$^\circ$ & --- & -3, +6 \\
W82.21 & 82209 km & 82190-82215 km &150.1$^\circ$ & 251.1$^\circ$ & --- & -2, -3, +6\\
W84.64 & 84644 km & 84625-84650 km & 145.4$^\circ$ & 195.5$^\circ$ & 235.5$^\circ$ & -2, +5, +6\\
W87.19 & 87189 km & 87175-87205 km & ---$^d$ & (182.2$^\circ$)$^e$ & 191.4$^\circ$ & -2, +5 \\ \hline
\end{tabular}}
\medskip
$^a$ Inferred resonance locations from \citet{Baillie11} Table 7\\
$^b$ Region over which wavelet power and phase are computed. \\
$^c$ $m$-values between -10 and +10 where the predicted $\delta \phi$ values are within $\pm30^\circ$ of the observed values. \\
$^d$ Phase difference not determined due to data gap \\
$^e$ Phase difference computed using wavevectors between $k=2\pi/(0.1$ km) and $k=2\pi/(2$ km) because otherwise combined peak average power is never above threshold. 
\end{table*}

\begin{figure*}
\centerline{\resizebox{6in}{!}{\includegraphics{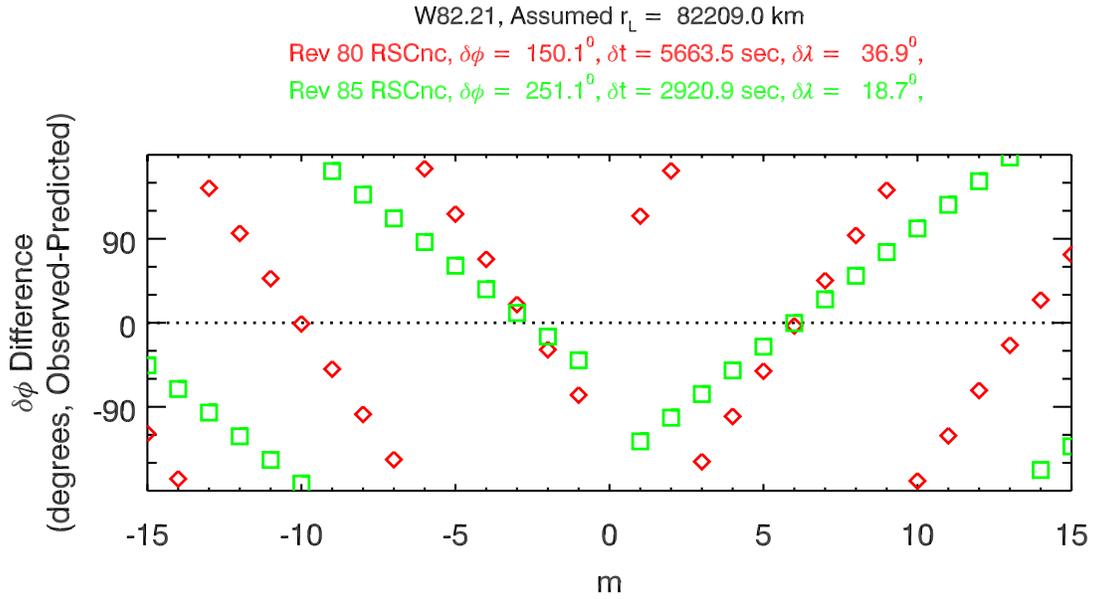}}}
\caption{Plot showing the difference between the observed and predicted values of $\delta \phi$ between the ingress and egress cuts of the RS Cancri occultations for the wave W82.21 as a function of the assumed $m$-number, given the stipulated $\delta \lambda$ and $\delta t$ values.  Different symbols correspond to different pairs of occultation cuts. Note that the residuals for both observations are close to zero when $m=-2, -3$ and $+6$, so these values of $m$ are the ones most consistent with the observed phase differences.}
\label{mphasecomp1}
\end{figure*}

\begin{figure*}
\centerline{\resizebox{6in}{!}{\includegraphics{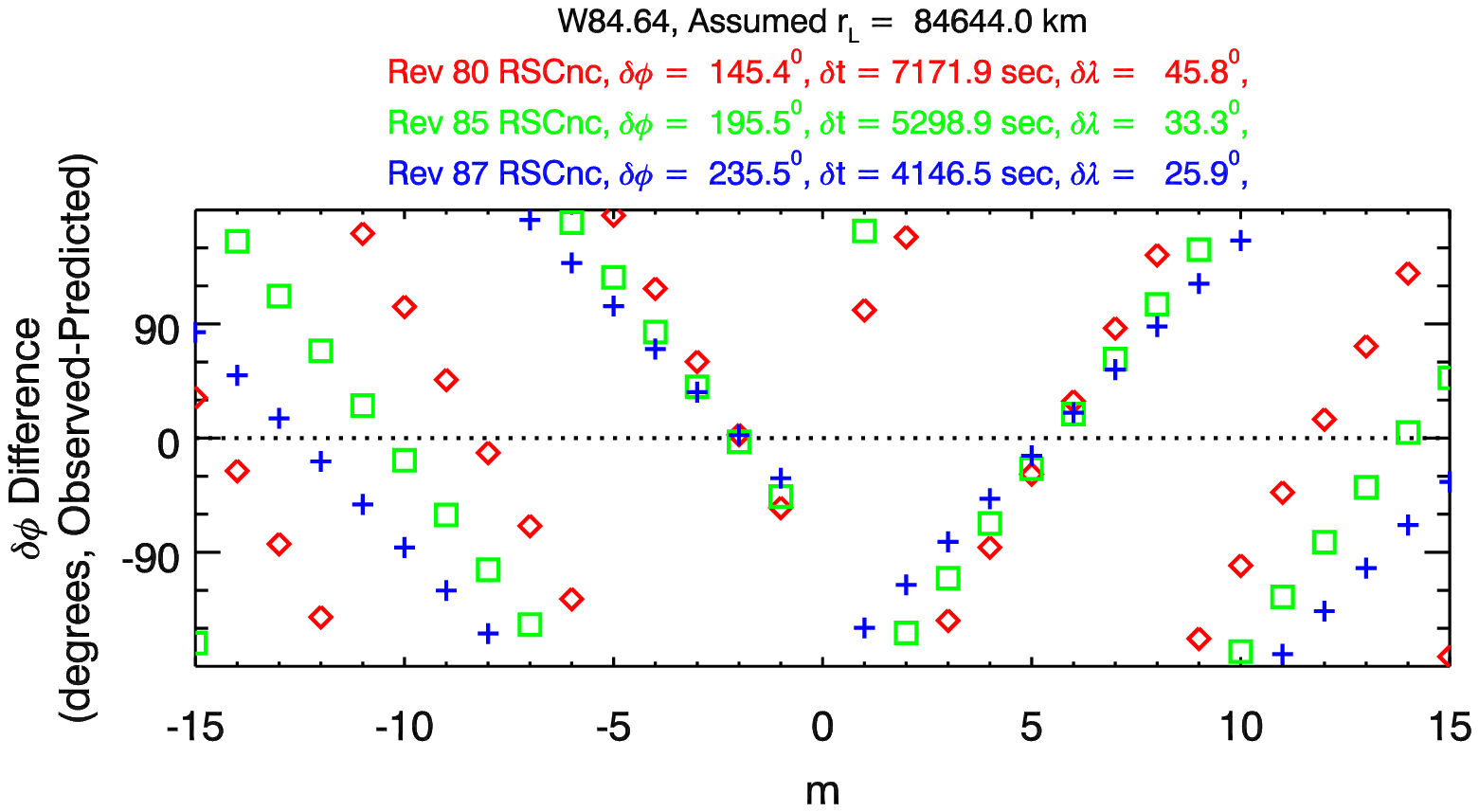}}}
\caption{Plot showing the difference between the observed and predicted values of $\delta \phi$ between the ingress and egress cuts of the RS Cancri occultations for the wave W84.64 as a function of the assumed $m$-number.  Note that the residuals for all three observations are close to zero when $m=-2, +5$ and $+6$, so these values of $m$ are the ones most consistent with the observed phase differences.}
\label{mphasecomp2}
\end{figure*}

In order to illustrate how quantitative estimates of  $\delta\phi$  can be used to constrain $m$-numbers and pattern speeds, let us examine the phase differences between the various RS Cancri ingress and egress profiles listed in {Table~\ref{waverscnc}}.  Note that the phase differences for the W82.00, W82.06 and W82.21 waves are indeed very similar for both the Rev 80 cuts (being around $\sim150^\circ$), and the Rev 85 cuts (where they are all $\sim250^\circ$). This is consistent with the above visual inspection of the profiles and supports our contention that these three waves must share a common $m$-number. However, with these measured values of $\delta\phi$, we can now also identify which values of $m$ and $\Omega_P$ come closest to predicting the observed phase differences.

For any given wave, the local mean motion $n$ and apsidal precession rate $\dot{\varpi}$ of the ring material are straightforward functions of the wave's resonant radius $r_L$ and Saturn's gravitational field \citep{MD99}. Hence, if we assume the gravitational field parameters given in \citet{Jacobson06}, we can use Equation~\ref{patspeed2} to compute the expected pattern speed $\Omega_P$ for any given value of $m$ at any of the relevant resonant radii. Furthermore, for a particular pair of occultation cuts, the difference in the observed ring longitudes $\delta \lambda$ and the difference in observation times  $\delta t$ can be obtained from Table~\ref{occtab1}. We can therefore use the computed pattern speed to calculate the expected phase difference between any two occultation cuts $\delta \phi({\rm predicted}) = |m|(\delta \lambda-\Omega_p\delta t)$ for each value of $m$ (cf. Eq.~\ref{dphi}).

Figures~\ref{mphasecomp1} and~\ref{mphasecomp2}  show the differences between the observed and predicted values of $\delta \phi$ for the waves W82.21 and W84.64 as a function of $m$.\footnote{Note that the time elapsed between the ingress and egress cuts can be as much as two hours. This is a non-trivial fraction of the 6.5-7 hour orbital period in this region and therefore cannot be ignored in the calculation of the predicted $\delta \phi$.}   For any given observation,  the difference $\delta \phi$(observed) - $\delta \phi$ (predicted) cycles repeatedly through 360$^\circ$ with increasing or decreasing $m$. Hence there are multiple possible  $m$-values that could be consistent with the single observed phase difference. However, different observations have different values of $\delta t$ and $\delta  \lambda$ and thus show different trends in these plots. Hence there are relatively few $m$-values that could be consistent with the two or three observations  illustrated in Figures~\ref{mphasecomp1} and~\ref{mphasecomp2}. If we conservatively assume a $\pm30^\circ$ uncertainty in the phase determinations (which is consistent with the results of the global analysis described below), then $m=$ -3 -2 and +6 are consistent with the RS Cancri observations of the W82.21 wave, and $m=$ -2, +5 and +6 are consistent with the observations of the W84.64. The other waves yield similar results, as shown in Table~\ref{waverscnc}. Note that the W82.00, W82.06 and W82.21 waves are consistent with the same set of $m$-values, as expected

\begin{figure*}[tbh]
\centerline{\resizebox{5in}{!}{\includegraphics{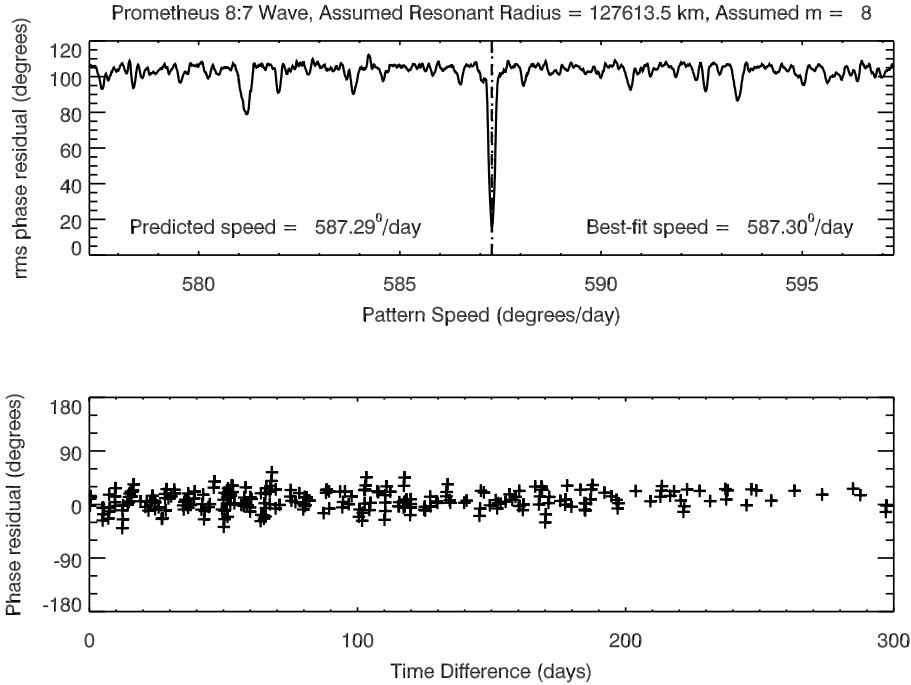}}}
\caption{A test of our pattern-speed determination algorithms using the Prometheus 8:7 wave in the inner A ring. The top panel shows the $rms$ phase difference residuals as a function of pattern speed assuming the pattern has an $m=8$, as expected for this wave. The dashed line marks the predicted pattern speed for this pattern, while the dotted line marks the pattern speed that gives the minimum variance (in this case, these two lines are almost on top of each other). The bottom panel shows the residuals in the phase differences from this best-fit solution as a function of time difference between the pairs of observations. The scatter in these data likely represents residual geometrical uncertainties in the various profiles.}
\label{prom87}
\end{figure*}

The small number of $m$-values that are consistent with this limited number of $\delta \phi$ estimates demonstrates that our measurements of $\delta \phi$ can constrain the symmetry properties and pattern speeds of these waves. However, it is also clear that we cannot uniquely determine the $m$-value for any of these waves with only the RS Cancri data. Thus the next step in this analysis is to extend this approach to include all pairs of occultation cuts listed in Table~\ref{occtab1}.

While the procedures described above allow us to calculate $\delta \phi$ for any possible pair of occultation cuts listed in  Table~\ref{occtab1}, in practice some occultation pairs do not yield reliable phase difference estimates. For some pairs, the average wavelet power for the two cuts never exceeds 0.9 of the peak power in each cut. This implies that the peak wave signal in the two cuts is coming from different parts of the wave (perhaps because they have different intrinsic resolutions), and so we do not consider such pairs.  We also exclude any occultation pairs which yield a $rms$ phase-difference scatter $\sigma_\phi$ greater than 20$^\circ$ because large values of $\sigma_\phi$ indicate that the wavelet was  unable to identify a consistent phase difference (perhaps due to small gaps or cosmic rays in one of the profiles). Finally, we only consider pairs of occultations where the time difference is less than 300 days. This prevents any aliasing that might occur due to the limited number of observations with larger time separations.  After applying these selection criteria, we have between 100 and 260 $\delta \phi$ estimates for each of the waves (see Table~\ref{wavetab2} and Table~\ref{phasetab}).
 
 As with the RS Cancri data described above, we constrain the symmetry properties and pattern speeds of each wave by computing expected values of $\delta \phi$ for different combinations of  $m$ and $\Omega_P$ and comparing these numbers with the observed $\delta \phi$ values. Specifically, we seek values of $m$ and $\Omega_P$ that minimize the $rms$ resdiuals of $\delta \phi$(observed) - $\delta \phi$ (predicted). In principle, we could just compute these $rms$ variations for each $m$ assuming $\Omega_P$ is given by Equation~\ref{patspeed2}, as we did for the RS Cancri data. However, given the uncertainties in the planet's gravitational field and the precise locations of the resonant radii for these waves, we will instead compute the $rms$ phase difference residuals over a finite range of pattern speeds surrounding the expected $\Omega_P$ corresponding to each $m$.  
 
 Since we are studying patterns that rotate around the planet at hundreds of degrees per day, and we are using occultations separated in time by up to 300 days, it is important to verify that our $\delta \phi$ calculations are sufficiently accurate to yield meaningful constraints on $m$ and $\Omega_p$. This is most easily done by first testing our methods on waves with known pattern speeds. We therefore applied the above algorithms to occultation profiles of the Prometheus 8:7 density wave in the outer A ring. The top panel in {Figure~\ref{prom87}} shows the $rms$ phase difference residual (observed-predicted) as a function of pattern speed, assuming the pattern has the expected $m=8$. Note the sharp dip at 587.30$^\circ$/day, which corresponds almost exactly to Prometheus' mean motion and the expected pattern speed for this wave.  Thus the pattern speed with the minimum variance in the phase residuals matches the expected perturbation frequency, as desired. Furthermore, no similar dip is seen when other values of $m$ are tried. These results demonstrate that the above procedures can indeed yield reliable estimates for the pattern speeds and $m$-numbers of even tightly-wound density waves.

The lower panel of Figure~\ref{prom87} shows the phase difference residuals as a function of time separation $\delta t$. These points are randomly scattered about zero, as desired, but with a $\pm 20^\circ$ scatter which probably reflects not only statistical uncertainties in the phase estimates themselves but also uncorrected systematic errors in the occultation geometries. As noted above, the latter can shift wave profiles relative to each other, which will produce errors in the phase differences that are hard to model. However, these data demonstrate that these systematic errors do not prevent us from determining the pattern speeds of spiral waves with wavelengths that are sufficiently long. More specifically, the above procedures should yield reliable results so long as  the uncertainties in the occultation geometry do not produce phase errors exceeding 90$^\circ$. As discussed in Section~\ref{waves} above, this should be the case for the six waves considered here.

The small dispersion in the phase difference residuals not only requires accurate phase difference measurements, the phase differences must also follow the predicted trend given by Eq.~\ref{dphi} over a sufficiently long period of time. For waves generated by satellites, the latter condition is equivalent to assuming the relevant moon has a constant mean motion, which is reasonable for most of Saturn's moons (the obvious exceptions being the co-orbitals Janus and Epimetheus, see Tiscareno  et al. 2006). \nocite{Tiscareno06} For waves generated by planetary normal modes, this condition requires assuming that the relevant oscillation has a constant frequency and a coherent phase over the time period spanned by the observations. This assumption not only turns out to be consistent with the observations (see below), it can also be justified {\it a priori} based on considerations of the waves' group velocity. 

Density waves in dense rings propagate away from the resonance at a finite speed, so
any shift in the frequency or phase of the perturbing potential generates discontinuities in the wave's profile \citep{Tiscareno06}. These features propagate through the rings at the group velocity $v_g \simeq \pi G \sigma_0/\kappa$, where $G$ is the universal gravitational constant, $\sigma_0$ the ring's average surface mass density, and $\kappa$ the radial epicyclic frequency \citep{Shu84}. For the relevant C-ring waves, $\sigma_0$  is between 1 and 10 g/cm$^2$ (Zebker et al. 1985, Hedman {et al.} 2011,  Section~\ref{discussion} below), and $v_g$ is between 0.4 and 2.5 km/year.
The relevant waves extend 10-20 km from the resonances and they do not exhibit any deviations from the expected smooth trends in their wavelengths. Thus, regardless of their origins, the frequencies and phases of the relevant perturbing potentials appear to be coherent over several years. Hence we are justified in assuming that the phase differences can be predicted using Eq.~\ref{dphi} for time intervals of one year or less.

\pagebreak

\section{Results}
\label{results}

\begin{figure*}
\centerline{\resizebox{5in}{!}{\includegraphics{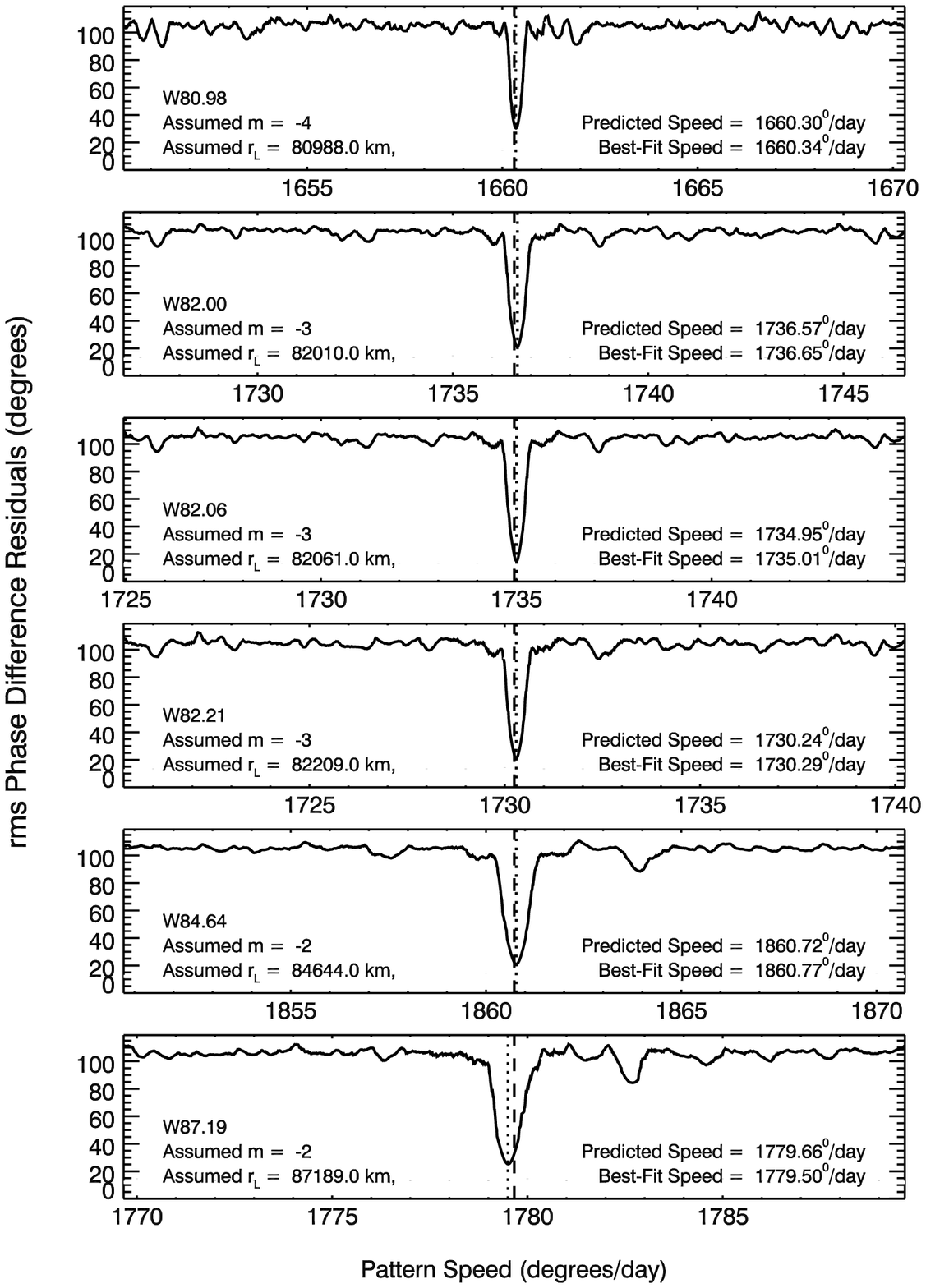}}}
\caption{Plots showing the $rms$ phase difference residuals as a function of pattern speed for each of the six waves, assuming the pattern has the indicated $m$-numbers. The dashed line marks the predicted pattern speed for this pattern at the resonant location provided by \citet{Baillie11}, while the dotted line is the pattern speed that gives the minimum variance in the residuals.}
\label{patspeeds}
\end{figure*}

\begin{figure*}
\centerline{\resizebox{5in}{!}{\includegraphics{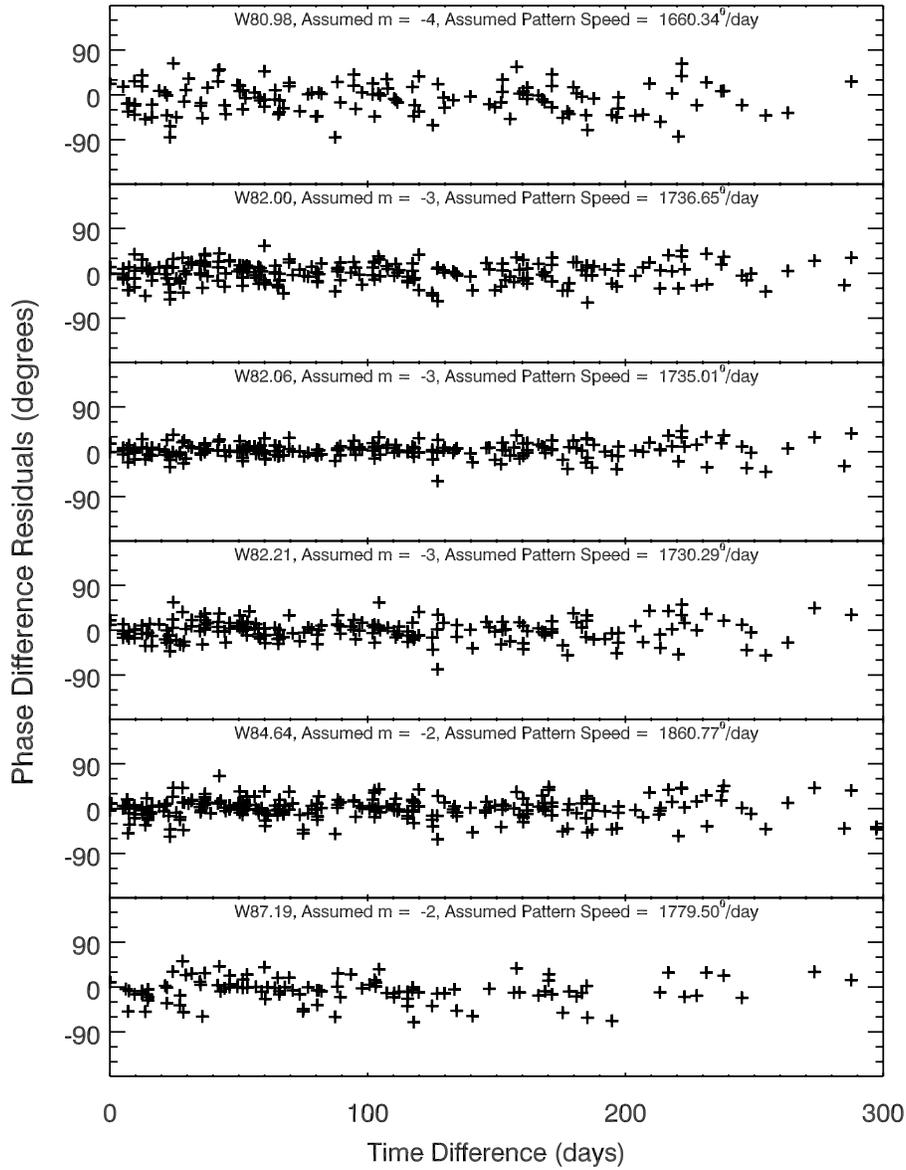}}}
\caption{Plots showing phase difference residuals (observed-predicted) for each of the six waves, assuming each pattern has the indicated $m$-number and pattern speed, which correspond to the best-fit value shown in Figure~\ref{patspeeds},}
\label{residuals}
\end{figure*}

\begin{table*}[tbp]
\caption{Results of the present analysis of the six waves}
\label{wavetab2}
\medskip
\centerline{\begin{tabular}{|c|c|c|c|c|c|c|c|} \hline
Wave & Resonant & Region & $N(\delta \phi)^c$ &  m & Pattern & Rotation  \\
 & Location$^a$ & Considered$^b$     & & & Speed$^d$ & Period \\ \hline
W80.98 & 80988 km &  80970-80995 km & 136 & -4 & 1660.3$^\circ$/day & 312.2 min \\
W82.00 & 82010 km & 81992-82012 km & 219 & -3 & 1736.6$^\circ$/day & 298.5 min \\
W82.06 & 82061 km & 82040-82065 km & 217 & -3 & 1735.0$^\circ$/day & 298.8 min \\
W82.21 & 82209 km & 82190-82215 km & 191 & -3 & 1730.3$^\circ$/day & 299.6 min\\
W84.64 & 84644 km & 84625-84650 km & 257 & -2 & 1860.8$^\circ$/day & 278.6 min\\
W87.19 & 87189$^e$ km & 87175-87205 km & 111 & -2 & 1779.5$^\circ$/day & 291.3 min\\ \hline
\end{tabular}}
\medskip
$^a$  Resonance locations from \citet{Baillie11} Table 7\\
$^b$ Region over which the wavelet phase and power are computed. \\
$^c$ Number of $\delta \phi$ estimates used in the fits. \\
$^d$ Best-fit value. A conservative estimate of the uncertainties in these numbers is 0.5$^\circ$/day, which would correspond to a $\sim180^\circ$ residual at 300 days. \\
$^e$ The true value may be closer to 87193 km (see text).
\end{table*}

We searched for the best-fitting combinations of $m$-numbers and pattern speeds for all six of the waves described in Section~\ref{waves}. Given that the wavelength of these waves increases with increasing radius, we expected these patterns would be generated by outer Lindblad resonances, and so the $m$-numbers of these waves would be negative. Also, previous studies of these waves by \citet{Rosen91} and \citet{Baillie11} suggested that if $|m|$ was above 5, the implied surface mass density of the C ring would be unreasonably high (see Section~\ref{discussion} below). Finally, the comparison of ingress and egress phases for the RS Cancri occultations in Table~\ref{waverscnc} suggests plausible values for $m$ of $-2, -3$ and $-4$.  Thus we expected that for these waves $m$ would lie between $-2$ and $-5$. Nevertheless, for the sake of completeness, we considered all $m$ values between $+10$ and $-10$, and searched for minima within 10$^\circ$/day of the expected pattern speed given by Equation~\ref{patspeed2}.\footnote{We also looked for patterns that would be produced by the higher-order Enceladus 5:1  ($m=2$, $\Omega_P=656.7^\circ$/day) and the Janus 5:2  ($m=3$, $\Omega_P=863.7^\circ$/day) inner vertical resonances that lay close to the W80.98 and W82.06 waves. No significant dips in $rms$ residuals were found.}

For each wave, we found a strong dip in the residual variance for only a single value of $m$. For wave W80.98, this dip occurred with $m=-4$; for each of waves W82.00, W82.06 and W82.21 it happened with $m=-3$; and for waves W84.64 and W87.19 it was $m=-2$. Figure~\ref{patspeeds} shows profiles of the $rms$ residuals versus pattern speed for all six waves, assuming the appropriate $m$-value. The resulting best-fitting pattern speeds can be compared with those calculated for each wave, using the resonance locations specified in Table 7 of \citet{Baillie11}. For all of the waves the minimum in residuals occurs very close to the expected pattern speed.  For five of the waves, the best-fit pattern speed is within 0.08$^\circ$/day of the value predicted using the \citet{Baillie11} resonance locations, while for the W87.19 wave the best-fit pattern speed is $0.16^\circ$/day lower than one would predict using the \citet{Baillie11} resonance location. These small offsets could represent small errors in either the assumed gravity field or the estimated resonant radii. For example, the offset between the best-fit and predicted pattern speeds for the W87.19 wave could be resolved if the real resonant radius $r_L$ lies just 4 km exterior to the \citet{Baillie11} estimate of 87,189 km. This wave is superimposed on a peak in optical depth (minimum in transparency), and the resulting variations in the surface mass density across the wave may have complicated earlier efforts to estimate the resonant radius. Indeed, looking at Figures~\ref{RCas106}, ~\ref{RSCnc80} and~\ref{RSCnc85}, the outer edge of this wave could easily fall somewhere around 87,193 km.

Figure~\ref{residuals} shows the individual phase difference residuals as a function of $\delta t$ for the best-fit solutions. These residuals typically have a spread of approximately $\pm$45$^\circ$, which is noticeably  larger than that found for the Prometheus 8:7 wave discussed above. This is not entirely surprising given the generally shorter wavelengths and more limited extents of the C-ring waves. Fewer cycles are visible in most occultation profiles of the C-ring waves than for the A-ring wave, so the estimates of the phase differences should be more uncertain. Furthermore, the maximum wavelength of the Prometheus 8:7 wave exceeds 10 kilometers, while the C-ring waves all have maximum wavelengths that are less than 5 kilometers. The estimated phases of the C-ring waves should therefore be more sensitive to small errors in the geometrical reconstructions. Indeed, if we do not include the small corrections to the occultation geometry provided by French {\it et al.} (in prep.), the $rms$ residuals of the best fit solutions increase by roughly a factor of two. However, even with this increased dispersion, the best-fit solution for each of the six waves remains close to the predicted pattern speed for the appropriate OLR. Hence, our identification of the best-fit $m$-numbers and pattern speeds appears to be robust against any residual uncertainties in the reconstruction geometry.

\section{Discussion}
\label{discussion}

{Table~\ref{wavetab2}} summarizes the results of this analysis, including the $m$-numbers, pattern speeds $\Omega_P$, and pattern rotation periods ($2\pi/\Omega_P$) for the six waves. We may conservatively estimate the uncertainties in the pattern speeds to be 0.5$^\circ$/day, since an error in the pattern speed of this magnitude would produce a trend in the residuals that would be clearly detectable in the data (the residuals would reach $\sim$180$^\circ$ at 300 days). This is also comparable to the widths of the minima in Figure~\ref{patspeeds}.  While the relatively tight distribution of the residuals already gives us some confidence that we have determined the correct values of $m$ and $\Omega_P$ for each of these waves, the following considerations lend additional support to these identifications:
\begin{itemize}
\item[1.]
Each of the pattern speeds is consistent with that calculated for an outer Lindblad resonance at that location, for the  appropriate $m$-number.
\item[2.]
Our full analysis indicates that the waves W82.00, W82.06 and W82.21 all have $m=-3$.  This is consistent with the similar observed phase shifts between ingress and egress observed for these three waves in the RS Cancri occulations discussed in Sections~\ref{rscnc} and \ref{pattern}. Recall that these arguments were  robust against uncertainties in the occultation geometry, and thus provide a useful check on the above analysis.
\item[3.]
The derived values of $m$ yield plausible values for the C-ring's surface mass density.
\end{itemize}

The last of these three points requires some explanation and qualification. In general, the variation of a wave's wavelength with radius can provide an estimate of the ring's average local surface mass density $\sigma_0$ and its mass extinction coefficient $\tau_n/\sigma_0$, via Eq.~(\ref{kr}). However, the estimated values of these parameters depend on $m$, so prior to this work researchers could only provide estimates of $\sigma_0/|m-1|$ for these six unidentified waves.  {Table~\ref{wavemass}} lists the estimates of $\sigma_0/|m-1|$ for all  six waves derived by \citet{Baillie11}, and the resulting estimates of $\sigma_0$ and $\tau_n/\sigma_0$ assuming the $m$-numbers derived  here.\footnote{Note that, for an OLR, $|m-1|$ becomes $|m|+1$.}  These values can be compared with estimates derived from other features in the rings, but in making these comparisons we must keep in mind that both $\sigma_0$ and $\tau_n/\sigma_0$ may vary with position across the ring. 

Studies of the five identifiable C-ring waves by \citet{Baillie11} yield extinction coefficients ranging between 0.13 and 0.36 cm$^2$/g, which are higher than our estimates. However, most of these identifiable waves occur in very different environments from those occupied by the six unidentified waves considered here. The two waves found near the Mimas 6:2 and Prometheus 4:2 resonances occur within a high-opacity plateau, while the waves associated with the Titan nodal resonance and the Mimas 4:1 resonance are found in rather low optical depth regions interior to 78,000 km. This leaves only the wave associated with the Atlas 2:1 ILR, which is within 500 km of W87.19 and not in a region of obviously elevated optical depth. Intriguingly, \citet{Baillie11} estimate that $\tau_n/\sigma_0 \simeq 0.19\pm0.04$ cm$^2$/g for the Atlas 2:1 wave, which is not too different from the value of  $0.11$ cm$^2$/g derived above for the nearby W87.19 wave.

For the waves between 79,000 km and 85,000 km there are no known density waves that can provide independent estimates for $\sigma_0$ or $\tau_n/\sigma_0$, but estimates of these parameters have been extracted from other types of observations. Based on the particle size distributions derived from Voyager RSS radio occultation data,  \citet{Zebker85} estimated the average surface mass density of the C ring  between 78,429 km and 84,462 km to be  $3.2\pm1.8$ g/cm$^2$. More recently,  \citet{Hedman11} found that variations in the wavelength of a vertical corrugation extending  between 78,000 and 84,000 km indicated that this part of the ring has a mass extinction coefficient of roughly 0.02 cm$^2$/g. Both of these numbers are reasonably consistent with the values of $\sigma_0$ and $\tau_n/\sigma_0$ given in Table~\ref{wavemass} for the inner five waves. We thus conclude that the $m$-values derived above do indeed yield plausible values for the mass density of the middle C ring. 
 
\begin{table*}[tbp]
\caption{Ring mass density and extinction coefficients}
\label{wavemass}
\medskip
\centerline{\begin{tabular}{|c|c|c|c|c|c|c|} \hline
Wave & m & $\sigma_0/|m-1|^a$ & $\tau_n^a$ & $(\tau_n/\sigma_0)|m-1|^a$ & $\sigma_0$ & $\tau_n/\sigma$ \\
 \hline
W80.98 &  -4 & 1.17 g/cm$^2$ & 0.13 & 0.11 cm$^2$/g &  5.85 g/cm$^2$ & 0.022 cm$^2$/g \\
W82.00 & -3 & 1.42 g/cm$^2$ & 0.14 & 0.10 cm$^2$/g &  5.68 g/cm$^2$ & 0.025 cm$^2$/g \\
W82.06 & -3 & 2.54 g/cm$^2$ & 0.28 & 0.11 cm$^2$/g &  10.16 g/cm$^2$ & 0.028 cm$^2$/g \\
W82.21 & -3 & 1.73 g/cm$^2$ & 0.13 & 0.08 cm$^2$/g &  6.92 g/cm$^2$ & 0.020 cm$^2$/g \\
W84.64 &  -2 & 1.35 g/cm$^2$ & 0.11 & 0.08 cm$^2$/g &  4.05 g/cm$^2$ & 0.027 cm$^2$/g \\
W87.19 &  -2 & 0.47 g/cm$^2$ & 0.15 & 0.33 cm$^2$/g &  1.41 g/cm$^2$ & 0.11 cm$^2$/g \\
 \hline
\end{tabular}}
\medskip
$^a$ from Table 7 of \citet{Baillie11}.
\end{table*}

Assuming the above $m$-numbers and pattern speeds are correct, we may now ask what could be producing these waves. They almost certainly cannot be produced by any of Saturn's moons, since such a moon would need to be orbiting just above Saturn's cloud tops, or within the tenuous D ring. Not only have no such moons been detected by Cassini, despite extensive imaging of this region, but any km-size icy or rocky objects unfortunate enough to find themselves here would be quickly torn apart by tidal forces. 

Instead, these waves may be generated by perturbations from oscillating normal modes within the planet. Indeed, using a particular model for Saturn's internal structure, \citet{MarleyPorco93} predicted pattern speeds for a number of normal modes that could produce several of the waves identified in the Voyager radio occultation \citep{Rosen91}. Those authors even suggested that W80.98 might be an $m=-4$ wave generated by a resonance with the $l=4$ sectoral (\ie\ $m=l$) $f$-mode in the planet (which they denote $4f$), and that either W82.06 or W82.20 might be generated by a similar $l=m=3$ planetary $f$-mode. Indeed, our fitted pattern speeds appear to be very consistent with Marley and Porco's predictions for the pattern speeds generated by normal mode oscillations inside the planet. Their predicted rotation period for the $4f$ mode of 310.6 minutes is quite close to the observed period of 312.2 minutes for the $m=-4$ wave W80.98. Their predicted $3f$ mode period of 296.4 minutes is close to the periods of the $m=-3$ waves W82.00, W82.06 and W82.21, which are $298.5-299.6$ minutes. Finally, the predicted $2f$ mode period of 286.4 minutes lies in between the observed periods of our two $m=-2$ waves W84.64 and W87.19. Hence, it seems quite likely that $f$-mode acoustic oscillations within the planet may be producing these six waves, in which case the precise pattern speeds presented here should be able to help constrain interior models for Saturn. (As but one example, we note that, within the range of pattern speeds calculated by \citet{Marley91} for various Saturn models, those which best fit the observed wave locations correspond to models with no interior differential rotation.) 

However, while each of the six waves has a pattern speed that fits with one of the normal modes studied by \citet{MarleyPorco93}, the number of waves associated with each normal mode is much harder to understand. \citet{MarleyPorco93} predicted that a {\em single} wave would be associated with each normal mode oscillation. Furthermore, only the $f$-modes with $m=l$ were predicted to have resonances in the C-ring (modes with $l>m$ or with internal radial nodes were predicted to have significantly faster pattern speeds and much smaller gravitational signatures). Thus they expected to find a single wave with $m=-2, -3$, $-4$, etc. Instead, we find three $m=-3$ waves in close proximity to each other, and two $m=-2$ waves with a larger separation. This suggests that Saturn's internal structure is more complex than Marley and Porco assumed. Thus future efforts to interpret these findings will likely need to consider the effects of differential rotation, compositional gradients, or a solid core.

At the same time, further investigations of the C-ring waves should provide additional constraints on potential models of Saturn's interior.  For example, once more accurate geometric reconstructions of the stellar occultations are available, it should be possible to examine the pattern speeds of shorter-wavelength waves, such as the waves designated ``d'' and ``h'' by \citet{Rosen91} or W85.67 and W83.63 by \citet{Colwell09}. These particular features could represent additional OLRs with $|m|>4$ predicted by \citet{MarleyPorco93}. Meanwhile, structures interior to 79,000 km could be generated by vertical resonances with a different class of planetary normal modes. There could even be additional $m=-4,-3$ or $-2$ patterns lurking within the rings that are either too subtle to detect in individual profiles or obscured by other ring features.

These  waves also potentially contain information about the amplitude of the planetary oscillations.  All else being equal, a larger-amplitude oscillation in the planet should generate larger fractional density variations in the corresponding density wave. This would imply that the $3f$ oscillations responsible for the W82.21 and W82.06 waves are larger than the one generating the W82.00 wave, and that the $2f$ oscillation responsible for W84.64 is larger than the one that gives rise to W87.19. However, the amplitude of the wave depends not only on the driving torque, but also on how quickly collisions among the ring particles dissipate coherent motions\citep{Shu84, Tiscareno07}. This complicates any effort to extract quantitative estimates of the perturbing force from the wave amplitude, especially for waves that exhibit large fractional density variations, like the ones considered here. 

Finally, the ring data can constrain how quickly the oscillations are generated and dissipated within Saturn's interior.  As mentioned in Section~\ref{pattern} above, the smooth trends in the waves' wavelength with radius indicate that each perturbing potential has maintained a coherent phase for several years. The small dispersion in the phase residuals illustrated in Figure~\ref{residuals} confirms this supposition, and furthermore we find no statistically significant trends in the dispersion with $\delta t$ out beeyond 300 days. If we assume that the planetary oscillations are stochastically excited, then this long coherence time implies a correspondingly high quality factor $Q$ for these oscillation modes. Indeed, given that the oscillation periods of the relevant modes in the planet's frame are $\sim 200$ minutes \citep{MarleyPorco93}, a coherence time in excess of 300 days would imply a $Q > 10,000$.  This number is comparable to limits computed from the expected tidal evolution of Saturn's moons over the age of the solar system \citep{Dermott88}, and is comparable to estimates of Jupiter's $Q$ derived from astrometric satellite data \citep{Lainey09}. However, recent analyses of astrometric measurements of Saturn's moons indicate that Saturn's  tidal $Q$ is between 1000 and 2000 \citep{Lainey12}. These apparently contradictory results might be reconciled if the lower-frequency tidal oscillations are predominantly dissipated by turbulent viscosity acting on inertial waves driven in the convective envelope, as proposed by \citet{OL04}. Further analyses of the full span of Cassini occultation data should provide novel constraints on dissipation rates within Saturn.


\section*{Acknowledgements}

The authors would like to acknowledge discussions with J. Fuller, N. Shabaltas and D. Lai at Cornell regarding the nature of planetary normal modes and their sensitivity to interior models, which greatly clarified our understanding of these phenomena. We also wish to thank M. Marley for his helpful review of this manuscript. A substantial improvement in our model fits (and reduction of residuals) was achieved  using unpublished corrections to the Cassini trajectory derived and provided by R.G. French. None of the data analyzed here would have existed without the dedicated work by many individuals on the Cassini project as well as the VIMS science and engineering teams. Financial support was provided by NASA.


\begin{table*}
\caption{Observed times (in seconds of ephemeris time, measured from the J2000 epoch) 
and inertial longitudes (measured relative to the longitude of ascending node on J2000) for the various occultation cuts through each wave.}
\label{occtab1}
\medskip
\begin{tabular}{|c|c|c||c|c|c|c|c|c|} \hline
 Star & Rev & i/e& W80.98 & W82.00 &W82.06 & W82.21 & W84.64 & W87.19 \\ \hline
R Hya&036&i&      220948260.&      220948093.&      220948085.&      220948061.&      220947679.&      220947295.\\
&&&         174.072$^\circ$&         175.060$^\circ$&         175.108$^\circ$&         175.249$^\circ$&         177.407$^\circ$&         179.453$^\circ$\\ \hline
$\alpha$ Aur&041&i&      227949007.&      227948789.&      227948779.&      227948748.&      227948261.&      227947783.\\
&&&         343.813$^\circ$&         345.077$^\circ$&         345.137$^\circ$&         345.316$^\circ$&         348.013$^\circ$&         350.513$^\circ$\\ \hline
$\gamma$ Cru&071&i&      266193414.&      266193267.&      266193260.&      266193238.&      266192889.&      266192521.\\
&&&         183.104$^\circ$&         183.241$^\circ$&         183.248$^\circ$&         183.267$^\circ$&         183.578$^\circ$&         183.886$^\circ$\\ \hline
$\gamma$ Cru&073&i&      267426088.&      267425942.&      267425935.&      267425913.&      267425565.&      267425200.\\
&&&         182.136$^\circ$&         182.281$^\circ$&         182.288$^\circ$&         182.309$^\circ$&         182.640$^\circ$&         182.969$^\circ$\\ \hline
$\gamma$ Cru&077&i&      269858216.&      269858071.&      269858064.&      269858043.&      269857699.&      269857336.\\
&&&         181.086$^\circ$&         181.240$^\circ$&         181.247$^\circ$&         181.269$^\circ$&         181.619$^\circ$&         181.965$^\circ$\\ \hline
$\gamma$ Cru&078&i&      270466692.&      270466548.&      270466541.&      270466520.&      270466176.&      270465814.\\
&&&         180.860$^\circ$&         181.015$^\circ$&         181.023$^\circ$&         181.045$^\circ$&         181.398$^\circ$&         181.747$^\circ$\\ \hline
$\beta$ Gru&078&i&&&&&      270512795.&      270512284.\\
&&&&&&&         302.800$^\circ$&         294.401$^\circ$\\ \hline
$\gamma$ Cru&079&i&      271045522.&      271045367.&      271045359.&      271045337.&      271044968.&      271044581.\\
&&&         179.175$^\circ$&         179.354$^\circ$&         179.363$^\circ$&         179.389$^\circ$&         179.795$^\circ$&         180.198$^\circ$\\ \hline
RS Cnc&080&i&      271872473.&      271872088.&      271872070.&      271872018.&      271871265.&\\
&&&          90.240$^\circ$&          87.858$^\circ$&          87.749$^\circ$&          87.431$^\circ$&          82.984$^\circ$&\\ \hline
RS Cnc&080&e&      271877226.&      271877611.&      271877629.&      271877681.&      271878434.&\\
&&&         121.515$^\circ$&         123.897$^\circ$&         124.005$^\circ$&         124.324$^\circ$&         128.770$^\circ$&\\ \hline
$\gamma$ Cru&081&i&      272320388.&      272320233.&      272320226.&      272320203.&      272319835.&      272319448.\\
&&&         178.322$^\circ$&         178.510$^\circ$&         178.519$^\circ$&         178.546$^\circ$&         178.974$^\circ$&         179.397$^\circ$\\ \hline
$\gamma$ Cru&082&i&      272956171.&      272956016.&      272956009.&      272955986.&      272955617.&      272955229.\\
&&&         177.862$^\circ$&         178.056$^\circ$&         178.065$^\circ$&         178.093$^\circ$&         178.532$^\circ$&         178.967$^\circ$\\ \hline
RS Cnc&085&i&&      275057262.&      275057224.&      275057119.&      275055932.&      275055073.\\
&&&&          97.236$^\circ$&          97.000$^\circ$&          96.337$^\circ$&          89.067$^\circ$&          84.114$^\circ$\\ \hline
RS Cnc&085&e&&      275059898.&      275059935.&      275060040.&      275061227.&      275062086.\\
&&&&         114.136$^\circ$&         114.371$^\circ$&         115.034$^\circ$&         122.303$^\circ$&         127.256$^\circ$\\ \hline
$\gamma$ Cru&086&i&      275503697.&      275503542.&      275503535.&      275503512.&&      275502756.\\
&&&         176.829$^\circ$&         177.033$^\circ$&         177.043$^\circ$&         177.073$^\circ$&&         177.995$^\circ$\\ \hline
RS Cnc&087&i&&&&&      276329999.&      276328974.\\
&&&&&&&          92.710$^\circ$&          86.701$^\circ$\\ \hline
RS Cnc&087&e&&&&&      276334140.&      276335165.\\
&&&&&&&         118.549$^\circ$&         124.557$^\circ$\\ \hline
$\gamma$ Cru&089&i&      277408751.&      277408596.&      277408589.&      277408566.&      277408199.&      277407813.\\
&&&         176.576$^\circ$&         176.781$^\circ$&         176.791$^\circ$&         176.821$^\circ$&         177.287$^\circ$&         177.749$^\circ$\\ \hline
$\gamma$ Cru&093&i&      280045204.&      280045028.&      280045020.&      280044994.&      280044576.&      280044136.\\
&&&         208.249$^\circ$&         208.061$^\circ$&         208.052$^\circ$&         208.024$^\circ$&         207.598$^\circ$&         207.175$^\circ$\\ \hline
$\gamma$ Cru&094&i&      280681410.&      280681250.&      280681242.&      280681218.&      280680836.&      280680433.\\
&&&         191.683$^\circ$&         191.696$^\circ$&         191.697$^\circ$&         191.699$^\circ$&         191.728$^\circ$&         191.758$^\circ$\\ \hline
$\gamma$ Cru&096&i&      282014259.&      282014112.&      282014104.&      282014083.&      282013732.&      282013362.\\
&&&         185.190$^\circ$&         185.280$^\circ$&         185.285$^\circ$&         185.298$^\circ$&         185.504$^\circ$&         185.708$^\circ$\\ \hline
$\gamma$ Cru&100&i&      285034037.&      285033857.&      285033848.&      285033822.&      285033398.&      285032956.\\
&&&         224.282$^\circ$&         223.835$^\circ$&         223.814$^\circ$&         223.749$^\circ$&         222.741$^\circ$&         221.750$^\circ$\\ \hline
$\gamma$ Cru&101&i&      285861190.&      285861011.&      285861002.&      285860976.&      285860552.&      285860110.\\
&&&         224.289$^\circ$&         223.842$^\circ$&         223.820$^\circ$&         223.755$^\circ$&         222.746$^\circ$&         221.755$^\circ$\\ \hline
$\gamma$ Cru&102&i&      286686360.&      286686182.&      286686173.&      286686147.&      286685725.&      286685285.\\
&&&         223.942$^\circ$&         223.500$^\circ$&         223.479$^\circ$&         223.415$^\circ$&         222.418$^\circ$&         221.438$^\circ$\\ \hline
$\beta$ Peg&104&i&      288914432.&      288914336.&      288914332.&      288914318.&      288914091.&\\
&&&         342.574$^\circ$&         343.021$^\circ$&         343.042$^\circ$&         343.107$^\circ$&         344.113$^\circ$&\\ \hline
R Cas&106&i&      291039691.&      291038969.&      291038939.&      291038853.&      291037730.&      291036830.\\
&&&          90.705$^\circ$&          86.728$^\circ$&          86.566$^\circ$&          86.097$^\circ$&          80.186$^\circ$&          75.723$^\circ$\\ \hline
$\alpha$ Sco&115&i&      302022977.&      302022638.&      302022621.&      302022571.&      302021771.&      302020939.\\
&&&         157.895$^\circ$&         158.409$^\circ$&         158.434$^\circ$&         158.508$^\circ$&         159.664$^\circ$&         160.797$^\circ$\\ \hline
\end{tabular}
\end{table*}

\begin{table}
\caption{Time, longitude and phase differences used to determine pattern speeds (full table included in on-line supplement)}
\label{phasetab}
\centerline{\begin{tabular}{|c|c|c|c|c|c|} \hline 
Wave  & Occultation Pair$^a$       &    $\delta t$ (days) &   $\delta\lambda$  (deg.) & $\delta\phi$ (deg.) &   $\sigma_\phi$ (deg.) \\ \hline
W82.00 & RSCnc085e-RSCnc085i  &        0.03051 &       16.9   &    250.6  &       3.0 \\
W82.06 & RSCnc085e-RSCnc085i   &       0.03137 &       17.4   &    249.1   &      5.7 \\
W82.21 & RSCnc085e-RSCnc085i&          0.03381 &       18.7   &    251.1   &      7.7 \\
W84.64 & RSCnc087e-RSCnc087i&          0.04799 &       25.9   &    235.5   &      4.2 \\
W80.98 & RSCnc080e-RSCnc080i&          0.05501 &       31.3   &    141.6   &      5.0 \\
W84.64 & RSCnc085e-RSCnc085i&          0.06133 &       33.3  &     195.5   &      8.9 \\
W82.00 & RSCnc080e-RSCnc080i&          0.06393  &      36.0   &    148.5   &      8.8 \\
W82.06 & RSCnc080e-RSCnc080i&          0.06434  &      36.3    &   148.1   &      6.1 \\
W82.21 & RSCnc080e-RSCnc080i&          0.06555  &      36.9     &  150.1   &      4.3 \\
W87.19 & RSCnc087e-RSCnc087i&          0.07165  &      37.9    &   191.4   &     12.1 \\
W84.64 & RSCnc080e-RSCnc080i&          0.08301  &      45.8    &   145.4   &      8.5 \\
W84.64 & betGru078i-gamCru078i&        0.53957    &   121.4    &    47.3     &    1.7 \\
W84.64 & gamCru081i-RSCnc080e&         5.10878  &      50.2  &     171.7   &      7.2 \\
W82.21 & gamCru081i-RSCnc080e&         5.12178   &     54.2   &    208.5    &     4.0 \\
W82.06 & gamCru081i-RSCnc080e &        5.12265    &    54.5   &    136.6     &    5.9\\
W82.00 & gamCru081i-RSCnc080e &        5.12294    &    54.6   &    110.6     &    2.3\\
W80.98 & gamCru081i-RSCnc080e &        5.12919    &    56.8  &      17.1     &   11.3 \\ \hline
\end{tabular}}
$^a$ Each occultation is designated by the sequence star-name, rev number and ingress/egress.
\end{table}

\end{document}